\documentclass[twocolumn]{IEEEtran}
\usepackage{fancyhdr}
\usepackage{cite}
\usepackage{amsmath}                
\usepackage{amsthm}                 
\usepackage{amssymb}
\usepackage{dsfont}
\usepackage{wasysym}
\usepackage{mathdots}
\usepackage[mathscr]{euscript}
\usepackage{graphicx}
\usepackage{eurosym}                
\usepackage{lscape}                 
\usepackage{tabu}
\usepackage{booktabs}
\usepackage{enumitem}
\usepackage{longtable}
\usepackage{multirow}
\usepackage{subfigure}
\usepackage{rotating}
\usepackage{color}                  
\usepackage[colorlinks=true, citecolor=blue, linkcolor=magenta]{hyperref}
\usepackage{stix} 			
\theoremstyle{definition}
\newtheorem{defn}{Definition}
\newtheorem{CaseStudy}{Case Study}

\theoremstyle{plain} 

\theoremstyle{remark} 

\usepackage{authblk}

\author[1,3]{Inas S. Khayal}
\author[2,3]{Amro M. Farid}
\affil[1]{The Dartmouth Institute, Geisel School of Medicine at Dartmouth, Lebanon, NH}
\affil[2]{Thayer School of Engineering at Dartmouth, Hanover, NH}
\affil[3]{Department of Computer Science, Dartmouth College, Hanover, NH}

\begin{document}

\title{A Dynamic System Model for Personalized Healthcare Delivery and Managed Individual Health Outcomes}


{}
\maketitle

\begin{abstract}
The current healthcare system is facing an unprecedented chronic disease burden. This paper develops a healthcare dynamic model for personalized healthcare delivery and managed individual health outcomes. It utilizes a hetero-functional graph theory rooted in Axiomatic Design for Large Flexible Engineering Systems and Petri nets. The dynamics of the model builds upon a recently developed systems architecture for healthcare delivery which bears several analogies to the architecture of mass-customized production systems. At its essence, the model consists of two synchronized Petri nets; one for the healthcare delivery system and another for individuals' health state evolution. The model is demonstrated on two clinical case studies; one acute and another chronic. Together, the case studies show that the model applies equally to the care of both acute and chronic conditions, transparently describes health outcomes and links them to the evolution of the healthcare delivery system and its associated costs.
\end{abstract}


\section{Introduction}
\vspace{-0.05in}
Growing healthcare costs have drawn significant attention to the healthcare delivery system and its fragile and fragmented nature.  Similarly, the growing burden of illness has also directed attention to addressing patients' health needs.  The consequences of the growing burden of illness compounded by an increasingly expensive healthcare delivery system places grave consequences on our economy and way of life.

Efforts to affect positive change requires an understanding of the complex dynamics of healthcare delivery systems and patients' health.  Most modeling focuses on either the 1.) \emph{healthcare delivery system} that renders patients without state as they are pushed and pulled through the system (e.g., a patient with an acute condition in an ER) or 2.) \emph{patient health} without any consideration of its interface with the healthcare delivery system.  In order to develop a dynamic system model of a personalized healthcare delivery system in which individual health outcomes are managed, these two processes need to be linked.    
 



\vspace{-0.2in}
\subsection{Dynamic Modeling of Healthcare Delivery Systems}
\vspace{-0.05in}
Efforts to quantitatively and dynamically model the healthcare delivery system originate from the field of production systems \cite{Zhong:2017:00}.  Production system modeling focuses on transporting operand-products one from location to another in the system.  The analogous extension to healthcare delivery system as a type of production system treats the patient as an operand-product as well.  In doing so, operand throughput and system efficiency in terms of cost and time can be quantitatively assessed and maximized.   
Consequently, a large body of healthcare literature summarizes safety \cite{saxena2017design, Carayon:2006:00}, resource management \cite{Fanti:2013:00}, capacity planning \cite{Garg:2012:00, Green:2005:00}, and scheduling of various types including for outpatients\cite{Wang:2012:00, Cayirli:2003:00,Rose:2011:00, kim2006stochastic}, clinicians\cite{Cheang:2003:00, Burke:2004:00, Ernst:2004:00}, operating room\cite{Cardoen:2010:00, Erdogan:2011:00, Guerriero:2011:00}), work flow processes \cite{Zhong:2016:00, Combi:2014:BPA,wang2012emergency, chan2015quick} and patient-flows\cite{Hall:2006:00}.

Healthcare is well-suited to apply production modeling methods for patients with acute conditions.  The focus is on addressing the urgency of the patient's health state by quickly transporting the patient through the healthcare delivery system before the patient falls into more serious diagnoses.  Such modeling assumes that the operand's state can be described by its position in the system in much the same way a product's state can be inferred based upon its relative position in a production system.  
  

Chronic conditions, however, have a much longer \emph{time-scale} and require a different approach.  
Moving patients through the healthcare delivery system efficiently (i.e., faster) does not address disease understanding or affect long-term disease course.  For example, addressing pain for a patient with a chronic condition (e.g., rheumatoid arthritis, fibromyalgia, cancer), may prevent recurring emergency department visits all-together and consequently modeling transportation during these visits is entirely superfluous.  Processing the acute visit and moving the patient through the ER faster, however,  does not address the need for long-term pain care.  Understanding health state and factors affecting health status, be they physical or social determinants of health (e.g., socioeconomic status, physical environment, social support) \cite{WHO:2008:00}, is critical when understanding an individual's interaction with the healthcare delivery system.  Given that chronic conditions now account for 78\% of healthcare delivery system expenditures\cite{Anderson2004b}, it is most important that modeling efforts focus on the most essential features of chronic care.

\vspace{-0.15in}
\subsection{Dynamic Modeling of Health}

Health has been modeled at many levels of granularity; from the cell level to the disease level.  The field of systems biology focuses on continuous-time modeling at a cellular scale of bio-physical-chemical processes\cite{Kitano2001, Kitano2002}.  In contrast, clinical medicine generally focuses on discrete modeling of diseases(e.g., diabetes \cite{Palmer:2004:00}, cancer \cite{Newton:2012:00}).  

 
\vspace{-0.15in}
\subsection{Paper Contribution}
This paper develops the dynamics for a system model for personalized healthcare delivery and managed individual health outcomes.  It utilizes a hetero-functional graph theory \cite{Schoonenberg:2017:ISC-BKR01} rooted in Axiomatic Design for Large Flexible Engineering Systems and Petri nets.  The dynamics of the model builds upon the developed systems architecture for healthcare delivery in \cite{Khayal:2017:HHS-J17} 
which bears several analogies to the architecture of mass-customized production systems.  At its essence, the model consists of two synchronized Petri nets; one for the healthcare delivery system and another for indviduals' health state evolution.    The model applies equally to the care of both acute and chronic conditions, transparently describes health outcomes and links them to the evolution of the healthcare delivery system and its associated costs.   

\vspace{-0.15in}
\subsection{Paper Outline}
The model's development rests upon a previously established architectural foundation\cite{Khayal:2017:HHS-J17}.  Section \ref{Sec:Background} presents the essential definitions and concepts from this foundation so that Section \ref{Sec:DynamicModel} may develop a conceptually consistent dynamic model.  Two illustrative examples are presented to demonstrate applicability to both acute care in Section \ref{Sec:ExampleAcute}, and chronic care in Section \ref{Sec:Example}.  Next, the discussion is presented in Section \ref{Sec:Discussion} and, finally, the conclusion in Section \ref{Sec:Conclusion}.  The work assumes prerequisite knowledge in model-based systems engineering \cite{Buede2011, Crawley2015, Weilkiens2011, Friedenthal2014}, graph theory \cite{Newman2010, VanSteen2010}, and discrete-event simulation \cite{Cassandras2008} which is otherwise gained from the cited texts.

\vspace{-0.1in}
\section{Background: Preliminaries}\label{Sec:Background}
The development of the dynamic model in Section \ref{Sec:DynamicModel} rests upon the recently developed architecture for personalized healthcare delivery and managed individual health outcomes.
That work drew upon a hetero-functional graph theory \cite{Schoonenberg:2017:ISC-BKR01} rooted in the Axiomatic Design for Large Flexible Engineering Systems and Petri nets.  The healthcare delivery system form is described by its resources in Section \ref{SystemForm}, and its system function is described by processes in Section \ref{SystemFunction}.  The processes are allocated to resources in the system concept as described by the system knowledge base in Section \ref{Concept-System}.
\vspace{-0.1in}
\subsection{System Form} \label{SystemForm}
The healthcare delivery system is composed of resources representing system form.  Four types of resources $\mathbb{R}=\mathbb{R}_F \cup \mathbb{R}_D \cup \mathbb{R}_M \cup \mathbb{R}_N$ have been defined \cite{Khayal:2017:HHS-J17}
:  
\begin{defn}
\textbf{Transformation Resource}\cite{Khayal:2017:HHS-J17}:  A resource $\mathbb{r}_F \in \mathbb{R}_F$ capable of a transformative effect on its operand (e.g. the health state of an individual). They are the set union of human and technical transformation resources, $\mathbb{R}_F =  R_F \cup \mathscr{R}_F$.  
\end{defn}
\begin{defn}
\textbf{Decision Resource}\cite{Khayal:2017:HHS-J17}:  A resource $\mathbb{r}_D \in \mathbb{R}_D$ capable of advising the operand, an individual, on how to proceed next with the healthcare delivery system.  
They are the set union of human and technical decision resources, $ \mathbb{R}_D=R_D \cup \mathscr{R}_D$.  
\end{defn}
\begin{defn}
\textbf{Measurement Resource}\cite{Khayal:2017:HHS-J17}: A resource $\mathbb{r}_M \in \mathbb{R}_M$ capable of measuring the operand: here the health state of an individual.  They are the set union of human and technical measurement resources, $\mathbb{R}_M =  R_M \cup \mathscr{R}_M$.  
\end{defn}
\begin{defn}
\textbf{Transportation Resource}\cite{Khayal:2017:HHS-J17}:   A resource $\mathbb{r}_N \in \mathbb{R}_N$ capable of transporting its operand:  the individual themself.  They are the set union of human and technical transportation resources, $\mathbb{R}_N =  R_N \cup \mathscr{R}_N$. \\
\end{defn}


\vspace{-0.2in}
\begin{defn} \label{Def:RBuffer}
\textbf{Buffer Resource}\cite{Khayal:2017:HHS-J17}:  A resource $r \in \mathbb{R}_B$, denoting specified locations as a set union of transformation, measurement and decision resources, 
where $\mathbb{R}_B=\mathbb{R}_F  \cup \mathbb{R}_D \cup \mathbb{R}_M$. 
\end{defn}


In the cases where a specific resource is capable of performing several processes, it is must be uniquely classified, such that if $r \in R$ can \emph{Transform}; then $r \in R_F$, then if $r \in R$ can \emph{Decide}; then $r \in R_D$, then if $r \in R$ can \emph{Measure}; then $r \in R_M$, otherwise $r \in R_N$ and $\mathscr{r} \in \mathscr{R_N}$ . 


\vspace{-0.1in}
\subsection{System Function} \label{SystemFunction}
The healthcare delivery system is composed of processes $P=P_F \cup P_D \cup P_M \cup P_N$ representing the system function.  Four types of processes have been defined \cite{Khayal:2017:HHS-J17}
:  
\begin{defn}
\textbf{Transformation Process}:  A \emph{physical} process $p_{F} \in P_{F}$ that transforms the operand: specifically the internal health state of the individual (i.e. treatment of condition, disease or disorder).  
\end{defn}
\begin{defn}
\textbf{Decision Process}:  A \emph{cyber-physical} process $p_{D} \in P_D$ occurring between a healthcare system resource and the operand:  the individual, that generates a decision on how to proceed next with the healthcare delivery system.  
\end{defn}
\begin{defn}
\textbf{Measurement Process}:  A \emph{cyber-physical} process $p_{M} \in P_M$ that converts a physical property of the operand into a cyber, informatic property to ascertain health state of the individual.  
\end{defn}
\begin{defn}
\textbf{Transportation Process}:  A \emph{physical} process  $p_{N} \in P_{N}$ that moves individuals between healthcare resources (e.g. bring individual to emergency department, move individual from operating to recovery room).\\

 \end{defn}
 \vspace{-0.2in}
\begin{defn}
\textbf{Non-Transportation Process}:  A combination of non-transportation processes representing transformation, decision and measurement process, $p_{B} \in P_{B}$ that is a set union of non-transportation processes, where $P_{B} = P_{F} \cup P_{D} \cup P_{M}$.
\end{defn}

\vspace{-0.2in}
\subsection{System Concept} 
\label{Concept-System}
The system concept is defined as an allocated architecture composed of a bipartite graph between system processes and resources, that can be mathematically described as \cite{Farid2007,Farid:2008:a1,Farid:2008:02,Farid:2015:IES-J19,Farid:2013:IEM-C30,Farid:2015:IEM-J23,Schoonenberg:2016:TES-JR06}, $P=J_S\odot \mathbb{R}$, 
where $J_S$ is the system knowledge base and $\odot$ is boolean multiplication.  
\begin{defn}
\textbf{System Knowledge Base} \cite{Farid2007,Farid:2008:a1,Farid:2008:02,Farid:2015:IES-J19,Farid:2013:IEM-C30,Farid:2015:IEM-J23,Schoonenberg:2016:TES-JR06}: A binary matrix $J_S$ of size $\sigma(P)\times\sigma(\mathbb{R})$ whose element $J_S(w,v)\in\{0,1\}$ is equal to one when event $e_{wv} \in {\cal E_S}$ (in the discrete event systems sense \cite{Cassandras2008}) exists as a system process $p_w \in P$ being executed by a resource $r_v \in \mathbb{R}$.   
\end{defn} 
\noindent The healthcare delivery system knowledge base $J_S$ represents the elemental capabilities that \emph{exist} within the system.  These capabilities may not always be \emph{available} and therefore such constraints can be described in a similar structure called the system events constraints matrix.  
\begin{defn}
\textbf{System Events Constraints Matrix}\cite{Farid2007,Farid:2008:a1,Farid:2008:02,Farid:2015:IES-J19,Farid:2013:IEM-C30,Farid:2015:IEM-J23,Schoonenberg:2016:TES-JR06}:  A binary matrix $K_S$ of size $\sigma(P)\times\sigma(\mathbb{R})$ whose element $K_S(w,v)\in\{0,1\}$ is equal to one when a constraint eliminates event $e_{wv}$ from the event set.  
\end{defn}
The construction of $J_S$ and $K_S$ allow for the construction of a \emph{system concept matrix} $A_S$ describing the independent actions defining the available capabilities in the system \cite{Farid2007,Farid:2008:a1,Farid:2008:02,Farid:2015:IES-J19,Farid:2013:IEM-C30,Farid:2015:IEM-J23,Schoonenberg:2016:TES-JR06}, $A_S=J_S \ominus K_S$,  
where $\ominus$ is Boolean subtraction. 
The enumeration of these independent actions defines the healthcare system's structural degrees of freedom. 
\begin{defn}
\textbf{Structural Degrees of Freedom}\cite{Farid2007,Farid:2008:a1,Farid:2008:02,Farid:2015:IES-J19,Farid:2013:IEM-C30,Farid:2015:IEM-J23,Schoonenberg:2016:TES-JR06}: The set of independent actions $\psi_i \in {\cal E_S}$ that completely defines the available processes in the system. It is given by:
\begin{align}\label{Eq:DOFS}
DOF_S=\sigma({\cal E}_S)&=\sum_w^{\sigma(P)}\sum_v^{\sigma(\mathbb{R})}A_S(w,v)
\end{align}
\end{defn}
\noindent From an architectural perspective, the structural degrees of freedom form the nodes of a hetero-functional network\cite{Farid:2016:IES-BC06, Farid:2015:IES-J19} that describes the structure of the healthcare delivery system.

It is often useful to vectorize the knowledge base, where the shorthand $()^V$ is used to replace vec(). A projection operator may be introduced to project the vectorized knowledge base onto a one's vector to eliminate sparsity. $\mathds{P}(A_S)^V=\mathds{1}^{\sigma{(\cal{E_S})}}$
such that \cite{Farid2007,Farid:2008:a1,Farid:2008:02,Farid:2015:IES-J19,Farid:2013:IEM-C30,Farid:2015:IEM-J23,Schoonenberg:2016:TES-JR06}: 
\begin{equation}\label{Eq:ProjMat}
\begin{array}{c}
\mathds{P} \\
\end{array}=
\left[\begin{array}{c}
e_{\psi_1}^{\sigma{(\cal{E_S})}}, \dots, e_{\psi_{\sigma{(\cal{E_S})}}}^{\sigma{(\cal{E_S})}} \\
\end{array}
\right]
\end{equation}
where $e_{\psi_1}^{\sigma{(\cal{E_S})}}$ is the $\psi_{i}^{th}$ elementary row vector corresponding to the first up to the last structural degree of freedom. 

\noindent In summary, the variables in the healthcare delivery structural system model are summarized in Table \ref{Ta:SummaryBkgd}. 
\vspace{-0.2in}
\begin{table}[h]
\caption{Healthcare Delivery Structural System Model Variables} \label{Ta:SummaryBkgd}
\vspace{-0.15in}
\begin{center}
\begin{tabular}{r l l}
\toprule
\multicolumn{1}{l}{\textbf{SYSTEM}} & \textbf{PERSONALIZED HEALTHCARE} \\
   & \textbf{DELIVERY SYSTEM} \\\toprule
\multicolumn{1}{l}{{\textbf{(A) System Form}}}  &  \\   
	Resources & buffer($R_B$)[transformation($R_F$) $\cup$  \\
	&decision($R_D$) $\cup$ measurement($R_M$)] $\cup$ \\
        & transportation($R_N$) \\\midrule
	Resource Classification  & transform>decide>measure>transportation  \\\midrule
\multicolumn{1}{l}{\textbf{(B) System Function}}  & \\
        Processes & transformation($P_F$) $\cup$ decision($P_D$) $\cup$ \\
	& measurement($P_M$) $\cup$ transportation($P_N$)\\\midrule					
\multicolumn{1}{l}{ \textbf{(C) System Context}} & \\  
   	System Knowledge Base & $J_S$ = $ \begin{bmatrix} J_{F} & 0 & 0 & 0\\  J_{FD} & J_{D} & 0 & 0\\  J_{FM}  & J_{DM} & J_{M} &  0\\  J_{FN} & J_{DN} & J_{MN} & J_{N}\\ \end{bmatrix}$   \\\midrule
  	System Constraint Matrix & $K_S$ = $ \begin{bmatrix} K_{F} & 0 & 0 & 0\\  K_{FD} & K_{D} & 0 & 0\\  K_{FM}  & K_{DM} & K_{M} &  0\\  K_{FN} & K_{DN} & K_{MN} & K_{N}\\ \end{bmatrix}$   \\\midrule
	System Availability Matrix & $A_S=J_S \ominus K_S$ \\\midrule
	Structural Degrees & $DOF_S=\displaystyle\sigma({\cal E}_S)=\sum_w^{\sigma(P)}\sum_v^{\sigma(\mathbb{R})}A_S(w,v)$ \\
	of Freedom & \\
\bottomrule
\end{tabular}
\end{center}
\end{table}


\section{Dynamic Model Development} \label{Sec:DynamicModel}
The structural model presented in the previous section provides a skeleton upon which to develop the dynamic model in this section.  Because healthcare delivery systems are spatially distributed and evolve with discrete-event dynamics, the dynamic model utilizes Petri nets \cite{Cassandras2008}.  Two types are needed.  The first is called the \emph{Healthcare Delivery System Petri Net}.  It describes the evolution of the system processes and resources of the healthcare delivery system in Section \ref{Sec:DynamicsHDS}.  Section \ref{Sec:ChronicCare} then refines this default model to the care of chronic conditions.  The second Petri net is called the \emph{Health Net}.  It describes the `clinical' health state evolution of individuals  in Section \ref{Sec:DynamicsIndividual}.  As discussed in detail previously \cite{Khayal:2017:HHS-J15}, although the human body's health state evolves continuously via biological processes, the practice of clinical medicine discretizes this evolution into discrete states so as to facilitate diagnosis and decision-making.   With these two Petri nets in place, their respective dynamics are synchronized in Section \ref{Sec:DynamicsSync}.  

\vspace{-0.2in}
\subsection{Healthcare Delivery System Dynamics} \label{Sec:DynamicsHDS}
The healthcare delivery system dynamics are described by a timed Petri net.  
\begin{defn}
\textbf{Healthcare Delivery System Petri Net:} A bipartite directed graph represented as a 6-tuple:  
\begin{equation}\label{Eq:SystemNet}
N= \{S, {\cal E}, {\cal M}, W, D, Q\} 
\end{equation}
where:  
\begin{itemize}
\item ${N}$ is the Healthcare Delivery System net.
\item $S$ is the set of places (or buffers) of size $\sigma{(\mathbb{R}_B)}$.
\item $\cal E$ is the set of transitions/events of size $\sigma{(\cal{E_S})}$.
\item $\cal M$ $\subseteq (S \times {\cal E}) \cup ({\cal E} \times S)$ is the set of arcs of size $\sigma{(\cal M)}$ from places to transitions and from transitions to places. 
\item $W:$ ${\cal M} \rightarrow \{0,1\}$ is the weighting function on arcs.
\item $D$ is the set of transition durations.
\item $Q$ is a discrete state marking vector of size $\left(\sigma(\mathbb{R}_B) + \sigma(\cal{E_S})\right) \times \mathrm{1} \in \mathds{N}^{{\sigma{(\mathbb{R}_B)}}+\sigma(\cal{E_S})}$.  
\end{itemize}
\end{defn}

In the model, there is exactly one \emph{place} for each healthcare system buffer.  As many healthcare systems may have hundreds or thousands of healthcare system buffers, it is often useful to form aggregated resources $\mathbb{\bar{R}}$\cite{Farid2007,Farid:2008:a1,Farid:2008:02,Farid:2015:IEM-J23,Khayal:2015:IES-J20}. 
\begin{equation}\label{Eq:DPAgg}
\mathbb{\bar{R}}=A_R\circledast \mathbb{R}
\end{equation}
where $\circledast$ is an aggregation operator and $A_R$ is an aggregation matrix\cite{Farid2007,Farid:2008:a1,Farid:2008:02,Farid:2015:IEM-J23,Khayal:2015:IES-J20} and $A_R(i,j)$=1 iff $\mathbb{R}_j \in \mathbb{\bar{R}}_i$.  For example, a human resource such as a surgeon must be aggregated with a technical resource such as an operating room in order to make a functional surgical theatre.  

In the model, there is exactly one \emph{transition} for every structural degree of freedom in the system.  This allows for all the capabilities of the healthcare delivery system to be potentially engaged by the patient population.  It is also important to note that the healthcare delivery system knowledge base can show process redundancies where a given process can be performed by multiple resources.   This critical distinction allows two different transitions to be fired and achieve the same process but engage entirely different resources at entirely different cost.   For example, the process of `perform skin suturing' performed by the resource `resident' vs. `plastic surgeon' have different costs associated with each transition.  

The (directed) \emph{arcs} of the Petri net graph and their weightings define the Petri net incidence matrix ${\cal M}$.    
\begin{defn}
\textbf{Petri Net Incidence Matrix} \cite{Farid2015}:  An incidence matrix $\cal M$ of size $\sigma{(\mathbb{R}_B)} \times \sigma{(\cal{E_S})}$ where: 
\begin{equation}
{\cal M} = {\cal M}^+ - {\cal M}^-
\end{equation}
where ${\cal M}^+ (y,\psi) =  $w$(\epsilon_{wv}, r_y) $ and ${\cal M}^-(y,\psi)= $w$(r_y, \epsilon_{wv})$ and $\psi$ is a unique index mapped from the ordered pair $(w,v)$.
\end{defn}
\noindent The incidence out and incidence in matrices (${\cal M}^-$ and ${\cal M}^+$) form the negative and positive components of the Petri net incidence matrix respectively.   The incidence out matrix may be calculated straightforwardly \cite{Schoonenberg:2016:TES-JR06}.  

\begin{align}\label{Eq:IncOutMat}
{\cal M}^-=\sum_{y1=1}^{\sigma(\mathbb{R}_{B})}&e_{y1}^{\sigma(\mathbb{R}_{B})}\left[\mathds{P}\left(X_{y_1}^-\right)^V\right]^T
\end{align} 
where,
\begin{align}
X_{y_1}^-=\left[\begin{array}{ccc}
\mathds{1}^{\sigma(P_B)}e_{y1}^{\sigma(\mathbb{R}_{B})T} & | & \mathbf{0}^{\sigma(P_B) \times \sigma({\mathbb{R}_N})} \\\hline
e_{y1}^{\sigma(\mathbb{R}_{B})} & \otimes \mathds{1}^{\sigma(\mathbb{R}_{B})} & \otimes \mathds{1}^{\sigma(\mathbb{R})T} \\
\end{array}\right]
\end{align} 
\noindent Equation \ref{Eq:IncOutMat} states that the incidence matrix is the linear superposition of $\sigma(\mathbb{R}_{B})$ matrices each associated with a given Petri net buffer $r_{y1}$.  For a given buffer $r_{y1}$, the outer product serves to link it to its associated structural degree of freedom or equivalently a Petri net transition.   Note that the matrix $X_{y_1}^-$ has the same size and structure as the system knowledge base $J_S$ and when projected by $\mathds{P}$ (in Equation \ref{Eq:ProjMat}) serves to select out the elements aligned with the structural degrees of freedom.  Finally, the $X_{y_1}^-$ matrix simply places filled elements at the structural degrees of freedom that 1.) occur at $r_{y1}$ and 2.) have $r_{y1}$ as its origin.  The incidence in matrix may be calculated analogously \cite{Schoonenberg:2016:TES-JR06}.  

\begin{align}
{\cal M}^+=\sum_{y2=1}^{\sigma(\mathbb{R}_{B})}&e_{y2}^{\sigma(\mathbb{R}_{B})}\left[\mathds{P}\left(X_{y_2}^+\right)^V\right]^T
\end{align} 
where,
\begin{align}
X_{y_2}^+=\left[\begin{array}{ccc}
\mathds{1}^{\sigma(P_B)}e_{y2}^{\sigma(\mathbb{R}_{B})T} & | &\mathbf{0}^{\sigma(P_B) \times \sigma({\mathbb{R}_N})}  \\\hline
\mathds{1}^{\sigma(\mathbb{R}_B)} & \otimes e_{y2}^{\sigma(\mathbb{R}_B)} &\otimes  \mathds{1}^{\sigma(\mathbb{R})T}\\
\end{array}\right]
\end{align} 

The Petri net structure leads directly to the definition of its timed discrete-event dynamics.
\begin{defn}
\textbf{Timed Petri Net (Discrete-Event) Dynamics} \cite{Farid2007,Farid:2008:a1,Farid:2008:02,Farid:2015:IES-J19,Farid:2013:IEM-C30,Farid:2015:IEM-J23,Schoonenberg:2016:TES-JR06}: 
Given a binary input firing vector $U^+[k]$ and a binary output firing vector $U^-[k]$ of size both of size $\sigma{(\cal{E}_S}) \times 1$, and the positive and negative components $\cal{M}^+$ and $\cal{M}^-$ of the Petri net incidence matrix of size  $\sigma{(\mathbb{R}_B)} \times \sigma{(\cal{E}_S)}$, the evolution of the marking vector $Q$ is given by the state transition function $\Phi_T(Q[k],U[k])$:
\begin{equation}
Q[k+1] = \Phi_T(Q[k],U^-[k], U^+[k])
\end{equation}
where $Q = [Q_S; Q_{\cal{E}_S}]$ and
\begin{align}
Q_{S}[k+1]     &= Q_{S}[k] & + & &{{\cal M}^+} U^+[k] & & - &&{{\cal M}^-} U^-[k] \\
Q_{\cal{E}_S}[k+1] &= Q_{\cal{E}_S}[k] & - & &U^+[k] && + &&U^-[k]   
\end{align}
\end{defn}

The state transition function breaks the discrete state $Q$ in two.  $Q_{S}$ tracks the locations of the tokens at the places $\mathbb{R}_B$ and $Q_{\cal{E}_S}$ tracks the locations of the tokens in the transitions $\cal{E}_S$ of the healthcare delivery system. The state transition function also distinguishes between input and output firing vectors so as to mark the entry and exit of tokens to and from transitions.  In practice, a scheduled event list is used to implement firing vectors and ensure the durations $D$ of each of the transitions.  
\begin{defn} \label{Def:ScheduledEventList}
\textbf{Scheduled Event List} \cite{Cassandras2008}: 
A tuple $\cal{S} =$ $(u_\psi[k], t_k)$ consisting of all elements $u_\psi[k]$ in firing vectors $U^-[k]$ and their associated times $t_k$.  For every element, $u_\psi^-[k] \in U^-[k]$, there exists another element $u_\psi^+[\kappa] \in U^+[\kappa]$ which occurs at time $t_\kappa$, $d_\psi$ time units later.  $t_\kappa=t_k + d_\psi$.  
\end{defn}

{\color{black}
Now that the dynamics have been defined, an operating cost function can be calculated based on the firing vectors, $U^+[k]$, in the scheduled event list.     \begin{defn}
\textbf{Cumulative Operating Cost Function:}  Operating costs incur as transitions fire, representing the execution of capabilities in the healthcare delivery system. 
Given a capability cost vector, $C$ of size $\sigma{(\cal{E}_S})\times1$, representing the cost for each capability and the input firing vector $U^+[k]$, the cumulative operating cost function, ${\cal C}$, is given by:    
\begin{align}
{\cal C}[t] = \sum_{k=1}^{t}C^TU^+[k],
\end{align} 
\end{defn} 
}

\vspace{-0.2in}
\subsection{The Chronic Condition Care Abstraction}\label{Sec:ChronicCare}
The healthcare delivery system model presented in the previous subsection considered all of its inherent capabilities and integrated them within a Petri net model.   Such an approach is considered sufficient for acute care and is demonstrated in Section \ref{Sec:ExampleAcute}.  For chronic care, however, several additional considerations are required.  First, because chronic conditions continue well beyond a single visit to a healthcare facility, a resource entitled `outside clinic must be included in the model.  Naturally, this will require the addition of transportation processes so as to enter and exit the clinic.   Next, transportation degrees of freedom within the clinic are assumed to have a negligible duration and are therefore eliminated.  $K_S$ is modified accordingly.  By Equation \ref{Eq:DOFS}, the number of structural degrees of freedom changes as well.  Consequently, a new projection operator $\mathds{P}_C$ must be calculated such that:
\begin{align}
{\mathds{P}_C} (J_S \ominus K_S)^V = \mathds{1}^{\sigma({\cal{E}_S})}
\end{align}
\noindent Finally, the resources within the clinic are aggregated by Equation \ref{Eq:DPAgg} so as to yield to $\mathbb{\bar{R}}$=\{healthcare clinic, outside clinic\}.  Consequently, the healthcare delivery system Petri net incidence out and incidence in matrices become:
\begin{align}
{\cal M}^-=A_R\sum_{y1=1}^{\sigma(\mathbb{R}_{B})}&e_{y1}^{\sigma(\mathbb{R}_{B})}\left[\mathds{P_C}\left(X_{y_1}^-\right)^V\right]^T
\end{align} 

\begin{align}
{\cal M}^+=A_R\sum_{y2=1}^{\sigma(\mathbb{R}_{B})}&e_{y2}^{\sigma(\mathbb{R}_{B})}\left[\mathds{P_C}\left(X_{y_2}^+\right)^V\right]^T
\end{align} 
This hierarchical aggregation implements the chronic condition care abstraction.  The focus now becomes the various forms of transformation, decision, and measurement processes that the patient receives rather than transportation and queuing within the clinic.  

\vspace{-0.2in}
\subsection{Health Net Dynamics} \label{Sec:DynamicsIndividual}
As mentioned previously, the Health Net
is introduced so as to represent the clinical health state of individuals.  
\begin{defn}\label{Def:HealthNet}
\textbf{Health Net}\cite{Khayal:2017:HHS-J17}
: Given an individual $l_i$, that is part of a population $L=\{l_1, ..., l_{\sigma(L)}\}$, the evolution of their clinical health state can be described as a fuzzy timed Petri net\cite{Pedrycz2003, Ding2005, Ding2006}:    
 \begin{equation}\label{Eq:HealthNet}
 {N}_{l_i}=\{S_{l_i},{\cal E}_{l_i},{\cal M}_{l_i}, W_{l_i}, D_{l_i}, Q_{l_i}\}
\end{equation}
where 
\begin{itemize}
\item ${N}_{l_i}$ is the health net. 
\item $S_{l_i}$ is the set of places describing a set of health states.
\item ${\cal E}_{l_i}$ is the set of transitions describing health events.
\item ${\cal M}_{l_i}$ $\subseteq (S_{l_i} \times {\cal E}_{l_i}) \cup ({\cal E}_{l_i} \times S_{l_i})$ is the set of arcs describing the relations of (health states to health events) or (health events to health states).
\item $W_{l_i}$ is the set of weights on the arcs describing the health transition probabilities for the arcs. 
\item $D_{l_i}$ is the set of transition durations.
\item $Q_{l_i}$ is the Petri net marking representing the likely presence of the set of health states as a discrete probabilistic state.
\end{itemize}
\end{defn}

The Petri net structure leads directly to the definition of its discrete-event dynamics.  
\begin{defn}\label{def:TPNDyn}
\textbf{Fuzzy Timed Petri Net (Discrete-Event) Dynamics}\cite{PopovaZeugmann:2013:00, Khayal:2017:HHS-J17}: Given a binary input firing vector $U_{l_i}^+[k]$ and a binary output firing vector $U_{l_i}^-[k]$ both of size $\sigma({\cal E}_{l_i}) \times 1$, and the positive and negative components ${\cal M}_{l_i}^+$ and ${\cal M}_{l_i}^-$ of the Petri net incidence matrix of size $\sigma(S_{l_i}) \times \sigma ({\cal E}_{l_i})$, the evolution of the marking vector $Q_{l_i}$ is given by the state transition function $\Phi(Q_{l_i}[k],U_{l_i}[k])$:
\begin{equation}\label{eq:Phi}
Q_{l_i}[k+1]=\Phi(Q_{l_i}[k],U_{l_i}^-[k], U_{l_i}^+[k])
\end{equation}
where $Q_{l_i}=[Q_{Sl_i}; Q_{{\cal E}l_i}]$ and 
\begin{align}\label{eq:Q_B}
Q_{Sl_i}[k+1]=&Q_{Sl_i}[k] &+& & M_{l_i}^+U_{l_i}^+[k] & &-& & M_{l_i}^-U_{l_i}^-[k]\\\label{eq:Q_E}
Q_{{E}l_i}[k+1]=&Q_{{E}l_i}[k] &-& & U_{l_i}^+[k] & &+& &U_{l_i}^-[k]
\end{align}
\end{defn}
\noindent $Q_{Sl_i}$ is introduced to probabilistically mark Petri net places whereas $Q_{{E}l_i}$ is introduced to mark the likelihood that a timed transition is currently firing.  
The transitions are fired based on a scheduled event list that combines the discrete events with a time interval as described in Definition \ref{Def:ScheduledEventList}. 

{\color{black}
Now that the health net and its dynamics are defined.  An individual's health outcome function can be calculated.  
\begin{defn}
\textbf{Health Outcome Function}:  An individual's health outcome is represented by the set of places describing health state, $S_{l_i}$.  Given a value vector that numerically represents health state, $V$ of size $\sigma(S_{l_i}) \times 1$, and the health state vector $Q_{Sl_i}[k]$, the health outcome function, ${\cal H}_{l_i}$, is given by:  
\begin{align}
{\cal H}_{l_i}[k] = V^T Q_{Sl_i}[k]
\end{align}
\end{defn} 
}
\vspace{-0.35in}
\subsection{Coordination of the Healthcare Delivery System Petri Net \& Individual Health Net Dynamics} \label{Sec:DynamicsSync}
As expected, the healthcare delivery system Petri net and the health net dynamics are inherently coupled.  Each transformation process in the healthcare delivery system induces its corresponding health event.  For each individual, $l_i$, this feasibility condition can be captured in a binary individual transformation feasibility matrix. 
\begin{defn}
\textbf{Individual Transformation Feasibility Matrix} $\Lambda_{F_i}$ \cite{Khayal:2017:HHS-J17, Farid2007,Farid:2008:a1,Farid:2008:02,Farid:2015:IES-J19,Farid:2013:IEM-C30,Farid:2015:IEM-J23,Schoonenberg:2016:TES-JR06}:  a binary matrix of size $\sigma({\cal E}_{l_i}) \times \sigma(P_F)$, where $\Lambda_{F_i}(x,j) = 1$ if transformational process $p_{F_j}$ realizes the health event $e_{xl_i}$.    
\end{defn}

An individual firing matrix is introduced to synchronize the healthcare delivery system Petri net firing vectors with those of the (individual) health nets.   
\begin{defn}
\textbf{Individual Health Firing Matrix} \cite{Farid:2016:TES-J27, Schoonenberg:2016:TES-JR06}:  A binary individual health firing matrix ${\mathcal U}[k]$ of size $\sigma({\cal E}_S) \times \sigma(L)$, whose element $u_{{l_i}{\psi,l}}[k] =1$ when the $k^{th}$ firing timing triggers an individual $l$ to take structural degree of freedom $\psi$ for action.  
\end{defn}

Consequently, the healthcare delivery system input firing vectors at a given moment $k$ become\cite{Schoonenberg:2016:TES-JR06} 
\begin{equation}
U^-={\mathcal U}\mathds{1}^{\sigma({L})}
\end{equation}
and each health net firing vector at a given moment $k$ becomes\cite{Schoonenberg:2016:TES-JR06} 
\begin{equation}
\Lambda_{F_i}^T \cdot U_{l_i} = {\cal A}_F \cdot {\cal U} \cdot e_{l_i}^{\sigma(L)T}
\end{equation}
and ${\cal A}_F$ serves to select out the structural degrees of freedom associated with transformation.

\section{Acute Care Illustrative Example} \label{Sec:ExampleAcute}
To demonstrate the model, an illustrative example of the acute care of an ACL injury and repair is chosen.  Section \ref{Sec:OrthoDescription} provides a narrative of the acute care episode.  Section \ref{Sec:OrthoSystem} then presents the healthcare delivery system model; first as a knowledge base, then in terms of a list of events, and finally as a Petri net.  Next, Section \ref{Sec:OrthoNet} presents the health net.  Finally, Section \ref{Sec:OrthoDyn} presents the coordination of the healthcare delivery system Petri net and individual health net dynamics.

\subsection{Description of Orthopedic Case}\label{Sec:OrthoDescription}
A typical example orthopedic case study of an ACL injury \& repair is described below; drawing from a textbook clinical case \cite{Guthrie2008}.
\vspace{-0.071in}
\begin{CaseStudy}\label{CS:Acute}
`Adam injured his left knee playing rugby when he fell forwards and sideways while the left foot remained fixed on the ground.  He felt immediate pain and was unable to continue with the game.  Pain and swelling increased over the next 2 hours. 
He was seen in an emergency department (ED) and X-rays were negative for fractures.  He was prescribed anti-inflammatories, given elbow crutches and advice on ice, rest and elevation.  A clinic appointment to see an orthopedic consultant was arranged.  

The orthopedic clinician (Ortho) evaluated the individual through a battery of special tests: anterior drawer test and valgus stress instability and active Lachman's test all of which were not conclusive due to pain and swelling.  The individual received an urgent MRI scan which showed a rupture of the left ACL and a medial collateral ligament tear.  Surgery was performed followed by an ACL post-operative rehabilitation protocol at physical therapy (PT)'. 
\end{CaseStudy}

\vspace{-0.25in}
\subsection{Modeling the Healthcare Delivery System of the Orthopedic Case}\label{Sec:OrthoSystem}
\vspace{-0.39in}
 \begin{figure}[h]
  \begin{center} 
   \includegraphics[scale=0.5, trim=2.1cm 5.9cm 1.81cm 6.1cm, clip=true]{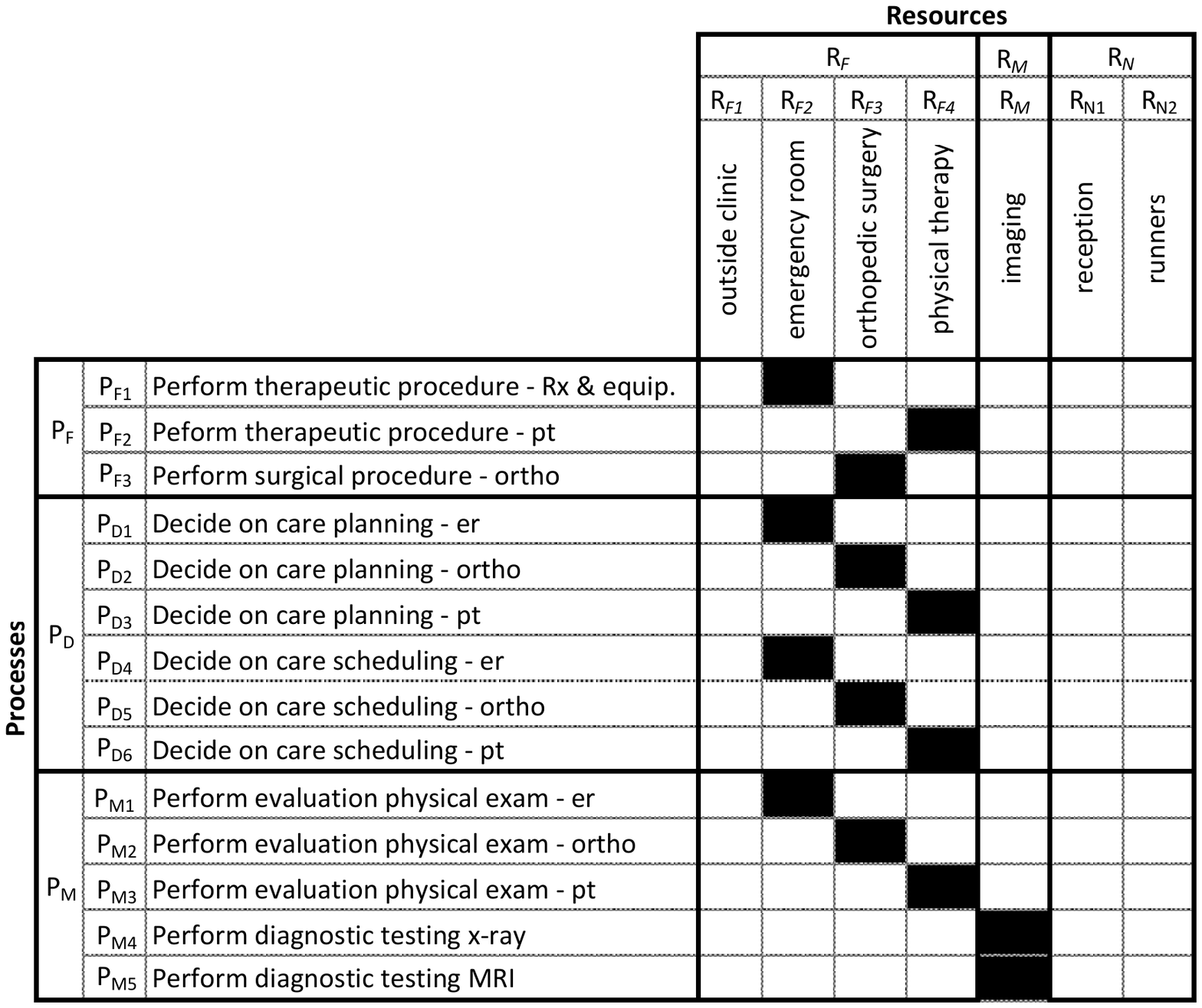} 
   \vspace{-0.3in}
    \caption{\textbf{Acute Care Healthcare Delivery System Knowledge Base} $J_{S}$.  Filled elements represent allocated processes to resources.   For graphical simplicity, the $J_{FN}$, $J_{DN}$, $J_{MN}$ and $J_{NN}$ have not been shown.}
   \label{Fig:JSOrtho}
   \end{center} 
   \vspace{-0.1in}
\end{figure}

\vspace{-0.1in}
The Case Study \ref{CS:Acute} text is interpreted so as to identify the healthcare delivery system processes and resources.   Each resource and process is then classified as either transformation, decision, measurement or transportation.  Additionally, an `outside clinic' resource is added to reflect the case's three clinical visits.  The resources and processes are used to construct the system knowledge base $J_{S}$ as shown in Figure \ref{Fig:JSOrtho}.  For graphical simplicity, $J_{FN}$, $J_{DN}$, $J_{MN}$ and $J_{NN}$ are not shown.   $J_{FN}$, $J_{DN}$, $J_{MN}$ are assumed to equal zero.  $J_{NN}$ introduces 20 transportation degrees of freedom within the clinic plus another 2 transportation degrees of freedom between the `reception' and `outside clinic'.  The text does not indicate any event constraints.  $K_S=0$.  The associated number of structural degrees of freedom is calculated from Equation \ref{Eq:DOFS}.  $DOF_S=36$.  
 \begin{figure}[h]
  \begin{center} 
   \includegraphics[width=2.5in]{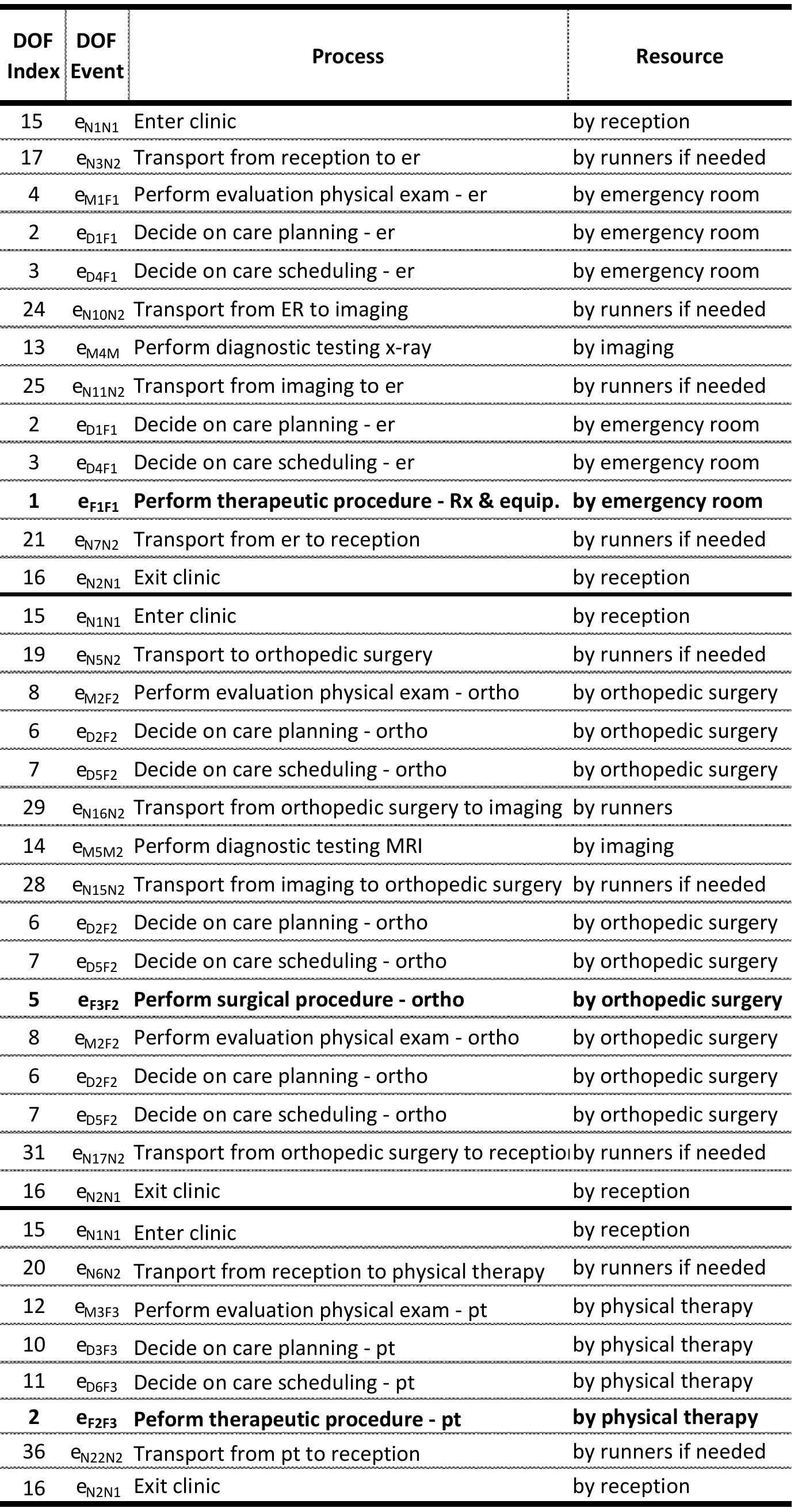} 
    \caption{\textbf{Acute Care Healthcare Delivery System Events} based on the Orthopedic Case narrative described in terms of the Healthcare Delivery System Events found in $J_S$.  The healthcare delivery transformational events are in bold. }
   \label{Fig:SystemeventsOrtho}
   \end{center}
\end{figure}

\vspace{-0.25in}
The Case Study \ref{CS:Acute} narrative is then rewritten as a string of healthcare delivery system events ${\cal E}_S$ as shown in Figure \ref{Fig:SystemeventsOrtho}.  Each event in ${\cal E}_S$ has a unique index and its associated combination of process and resource.  The transformational events are highlighted in bold.  These events are effectively an {\color{black}\emph{untimed}} scheduled events list and are used to generate the healthcare delivery system Petri net firing vectors.

 \begin{figure}[h]
  \begin{center} 
   \includegraphics[width=3.25in, trim=0.009cm 0.0095cm 0.062cm 0.0532cm, clip=true]{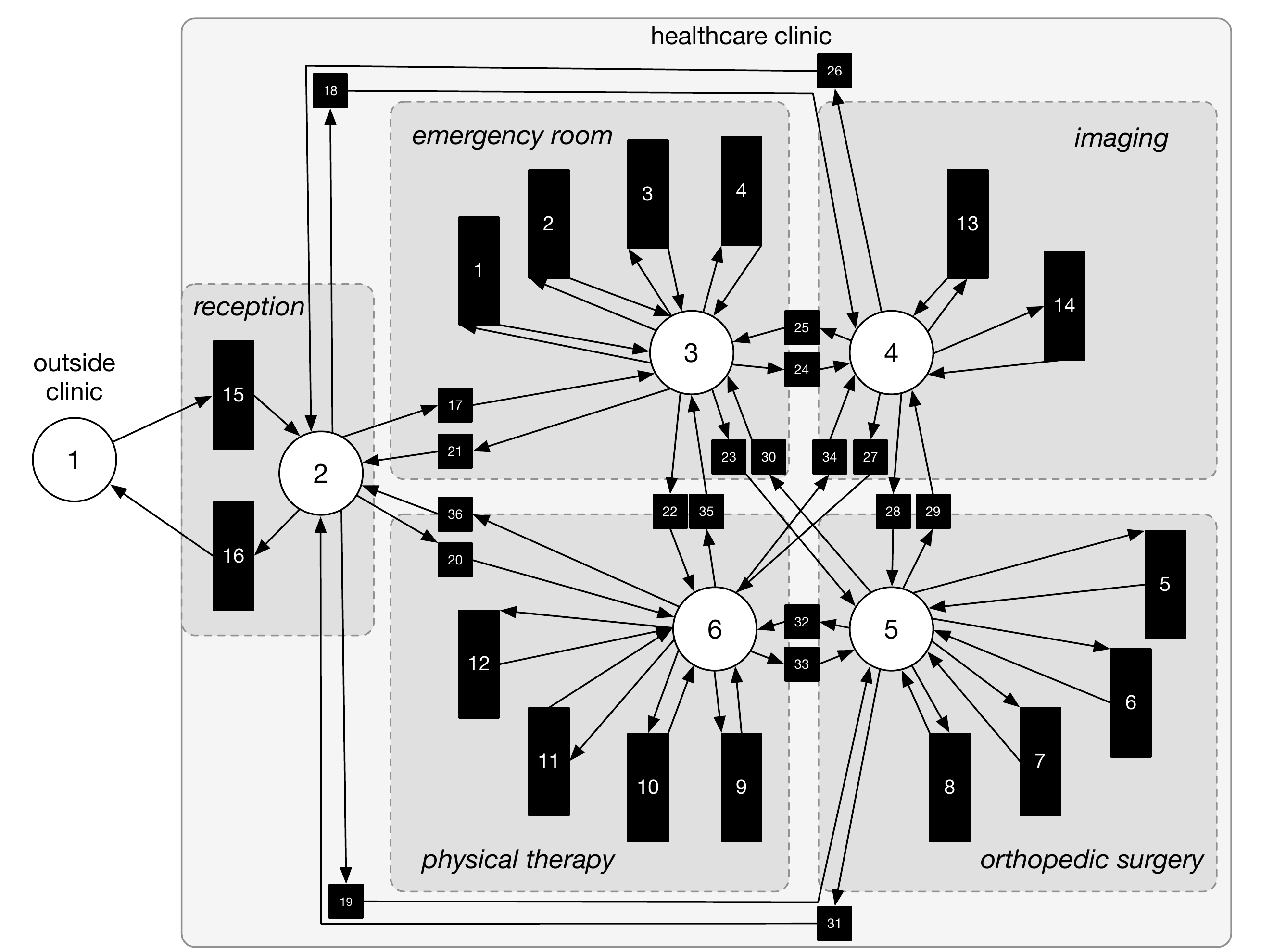} 
\vspace{-0.1in}
    \caption{\textbf{Acute Care Healthcare Delivery System Petri net}.  The places (\emph{circles}), represent buffer resources=\{outside clinic, reception, emergency room, imaging, physical therapy, orthopedic surgery\}.  The transitions (\emph{rectangles}), are numbered based on the healthcare delivery system structural degrees of freedom index  $\psi_i$.}
   \label{Fig:HDSnetOrtho}
   \end{center}
\vspace{-0.2in}
\end{figure}

The healthcare delivery system Petri net model is constructed as shown as in Figure \ref{Fig:HDSnetOrtho}.  A single Petri net place is shown for each of the 6 buffer resources.   A single transition is shown for each of the 36 structural DOF.   The places and transitions have been graphically situated to reflect their `physical' location in the healthcare clinic.

\vspace{-0.2in}
\subsection{Modeling the Individual Health Net Episode of the Orthopedic Case}\label{Sec:OrthoNet}

\begin{figure}[h]
\vspace{-0.2in}
  \begin{center}
   \includegraphics[scale=.34,trim=0.06cm 0.25cm 0.3cm 0.2cm, clip=true]{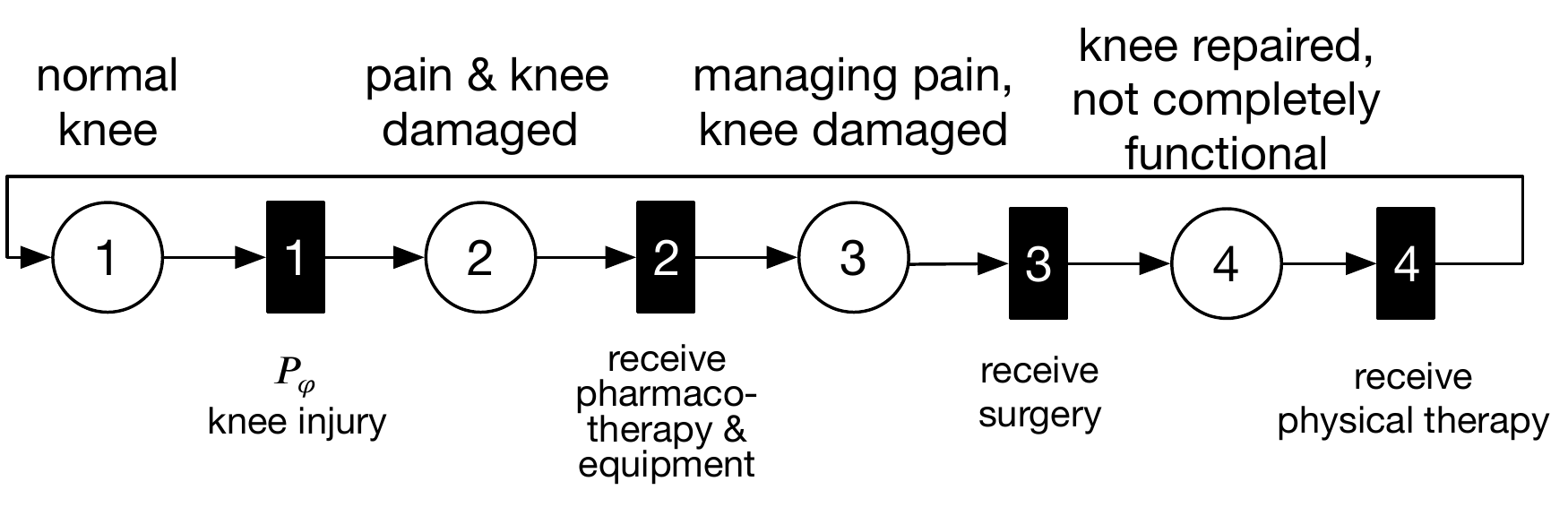}
    \vspace{-0.1in}
   \caption{\textbf{Acute Care Individual Health Net}: visualizes the health state at the places \emph{(circles)} and the health events, causing the changes in health state, at the transitions \emph{(rectangles)}.}
   \label{Fig:OrthopedicNet}
   \end{center}
\vspace{-0.25in}
\end{figure}

The Case Study \ref{CS:Acute} narrative and its associated healthcare delivery system events now serve to determine the individual health net $N_l$ as shown in Figure \ref{Fig:OrthopedicNet}.  

The health net shows the individual's \emph{health states} at the places (circles) and the individual's \emph{health state transformations} or \emph{health events} at the transitions (rectangles).  These occur due to healthcare delivery system events $P_F$ or stochastic human processes $P_\varphi$.   As is common with acute conditions, there is a serial progression of events which when successful return the patient back to a normal health state.  

\subsection{Modeling the Coordination of the Healthcare Delivery System Petri Net \& Individual Health Net of the Orthopedic Case}
\label{Sec:OrthoDyn}
\begin{figure}[h]
  \begin{center} 
  \vspace{-0.1in}
   \includegraphics[scale=0.4, trim=1.38cm 6.86cm 2.3cm 2.11cm, clip=true]{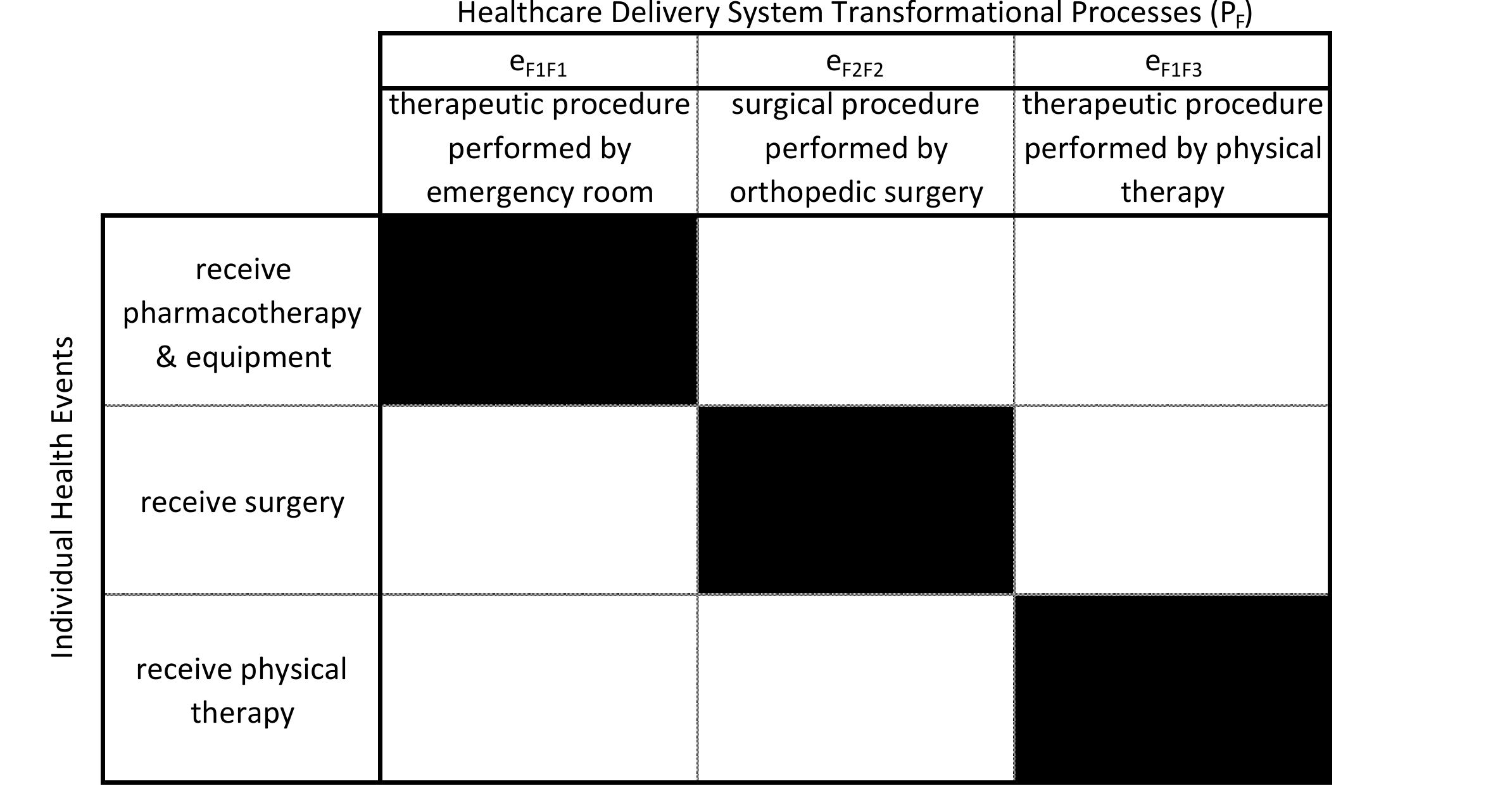} 
   \vspace{-0.26in}
    \caption{\textbf{Acute Care Individual Transformation Feasibility Matrix} $\Lambda_{F}$.}
   \label{Fig:Ex1FeasibilityMatrix}
   \end{center}
   \vspace{-0.2in}
\end{figure}

To complete the model, the healthcare delivery system Petri net and individual health net must be coordinated.  The individual transformation feasibility matrix, shown in Figure 
\ref{Fig:Ex1FeasibilityMatrix}, is constructed by linking the individual health net transitions (i.e. health events) to the corresponding healthcare delivery system transformational events (i.e. transformation process $P_F$). 

\begin{figure}[]
  \begin{center} 
   \includegraphics[width=3.5in]{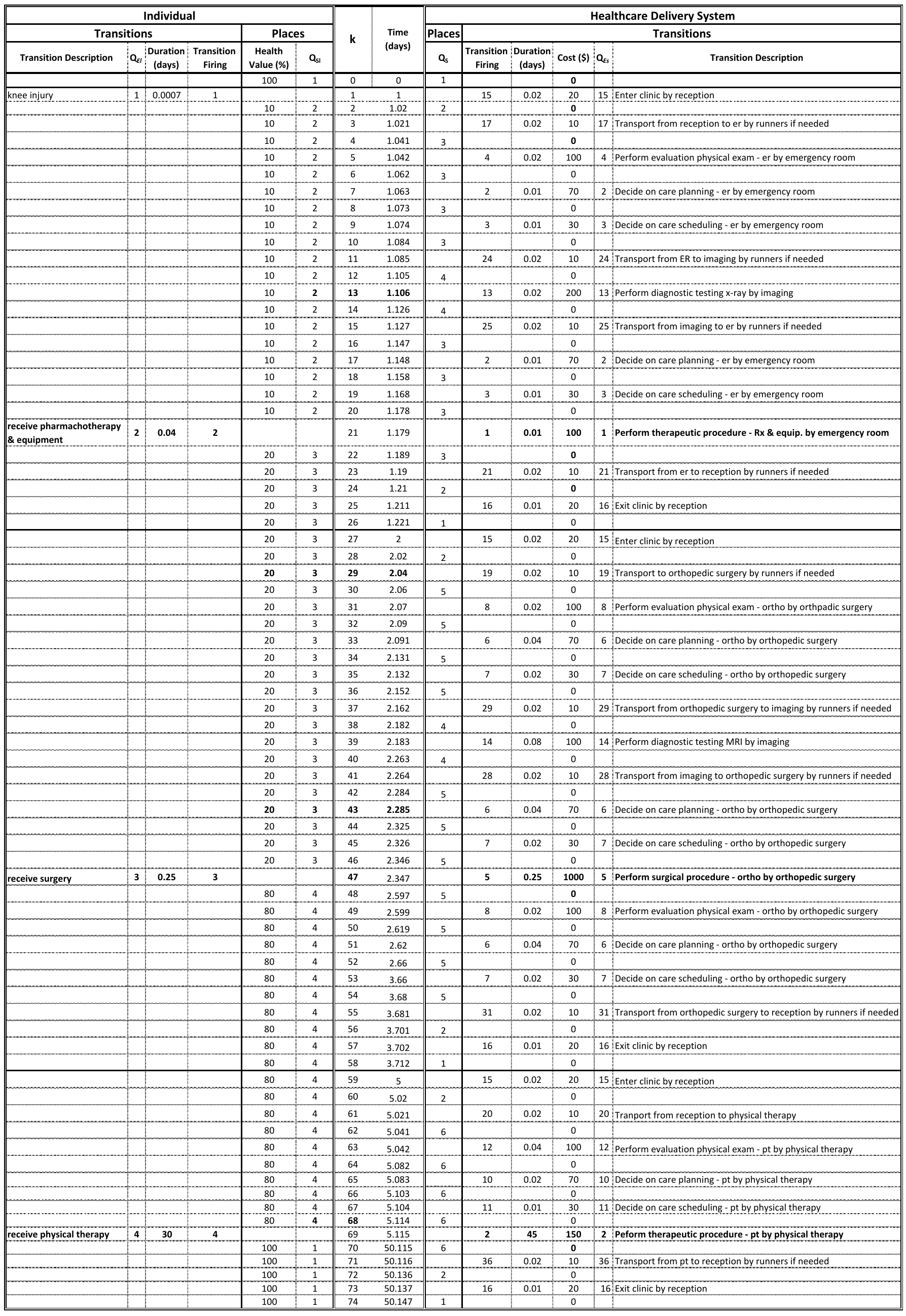} 
  \vspace{-0.3in}
    \caption{\textbf{Acute Care Dynamics of the Two Petri nets}:  the healthcare delivery system Petri net and the individual net are now synchronized.  The scheduled events of each Petri net are shown side by side.}
   \label{Fig:DynamicTableOrtho}
   \end{center}
   \vspace{-0.295in}
\end{figure}

The synchronized dynamics of the healthcare delivery system Petri net and the individual net are shown in Figure \ref{Fig:DynamicTableOrtho} as two scheduled event lists side by side.

{\color{black} Finally, the healthcare delivery system operating cost and the individual health outcome dynamics for this Orthopedic acute case can be shown over time (k) in Figure \ref{Fig:Cost_HealthOutcome}.
 \begin{figure}[h]
  \begin{center} 
   \includegraphics[width=3.in]{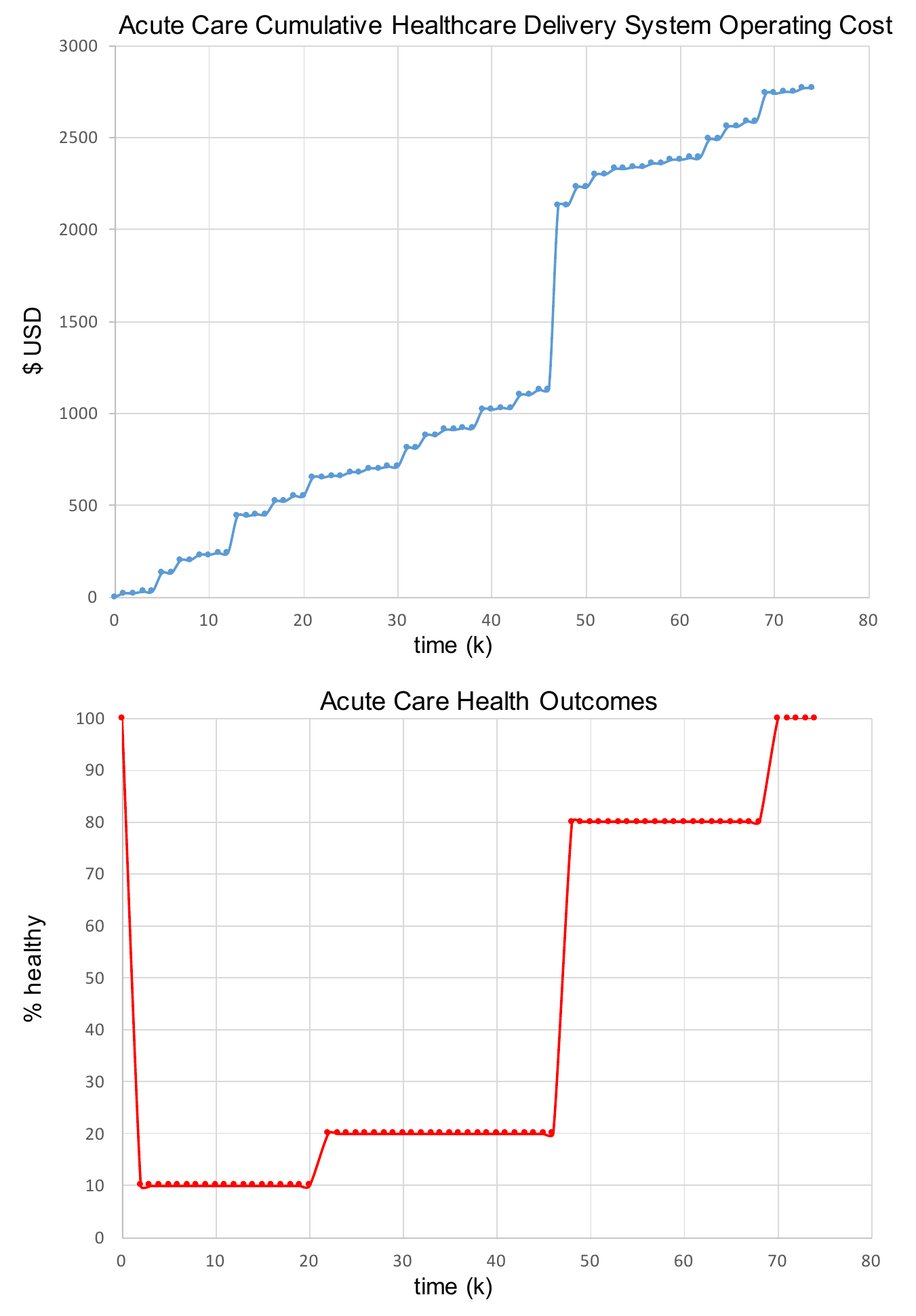} 
      \vspace{-0.17in}
        \caption{\textbf{Acute Care Healthcare Delivery System Operating Cost in US dollars (top-blue) and Individual Health Outcome as \% healthy (bottom-red) over time k. }
        }
   \label{Fig:Cost_HealthOutcome}
   \end{center}
\vspace{-0.3in}
\end{figure}
}

\vspace{-0.2in}
\section{Chronic Care Illustrative Example} \label{Sec:Example}
In contrast to the previous example, a neuro-oncology case study is chosen to demonstrate the model's applicability to chronic conditions.  Section \ref{Sec:Description} presents the full narrative of the case as presented by Park et. al. \cite{Park2005}.  Section \ref{Sec:ModelHDS} then presents the healthcare delivery system model; first as a knowledge base, then in terms of a list of events, and finally as a Petri net.  Next, Section \ref{Sec:Modelindividual} presents the health net.  Finally, Section \ref{Sec:Dyn} presents the coordination of the healthcare delivery system Petri net and individual health net dynamics.

\vspace{-0.25in}
\subsection{Description of Neuro-Oncology Case} \label{Sec:Description}
\begin{CaseStudy}\label{CS:Chronic}`The patient was a 32-year-old, right-handed woman without significant past medical history who presented for evaluation of headaches and intermittent short-term memory loss. She also
reported mild nausea, but was otherwise asymptomatic. Her headaches had begun approximately 3 months before her presentation.  On neurologic examination, no deficit was appreciated.  She underwent a head CT with the finding of a large, poorly enhancing right occipital-parietal mass that appeared to be located within the lateral ventricle. Subsequent MR imaging scanning confirmed the location of the tumor in the trigone with local expansion of the ventricle.  Similar to the CT, minimal enhancement was noted. On the  basis of the imaging characteristics, a low-grade astrocytoma was felt to be the most likely diagnosis. In light of the size of the tumor and its location, a parietooccipital surgical approach was performed. The tumor appeared grayish and was predominantly firm, necessitating piecemeal removal. The tumor was not particularly vascular, and a distinct plane between tumor and ependyma was identified, but there were several areas where the tumor appeared to infiltrate into adjacent brain parenchyma. Frozen-section pathologic analysis was described as abnormal and cellular but was not specifically diagnostic.  Tumor resection was continued until a near-total removal was accomplished. Postoperatively, the patient remained without neurologic deficit. Follow-up MR imaging showed near-total removal of the tumor.

Histologic sections were examined by light microscopy. The neoplasm was hypercellular with necrosis and endothelial cell proliferation, hallmarks of GBM, World Health Organization (WHO) classification grade IV (9, 10). Gemistocytes were distributed throughout the neoplasm with rare mitoses. These cells had hyaline, eosinophilic cytoplasm, and eccentric, hyperchromatic nuclei, some of which were large and pleomorphic. Immunostains for glial fibrillary acidic protein (GFAP), S-100 protein, vimentin, and neurofilament were positive, whereas immunostains for muscle-specific actin, alpha smooth muscle, and synaptophysin were negative. MIB-1 was positive, with a low proliferation index. The morphology and GFAP positivity suggested a diagnosis of diffuse gemistocytic astrocytoma; however, the tumor necrosis and microvascular proliferation raised the tumor grade to grade IV GBM. Because of the unusual radiographic appearance and location, the histologic slides were also reviewed by physicians at the Armed Forces Institute of Pathology (Bethesda, MD), who confirmed the initial diagnosis. Whole-brain radiation and chemotherapy were subsequently instituted. At 2-year follow-up, the patient complained of headaches but remained neurologically intact with no evidence of tumor progression on MR imaging.' \cite{Park2005}
\end{CaseStudy}

\vspace{-0.23in}
\subsection{Modeling the Healthcare Delivery System of the Neuro-Oncology Case}\label{Sec:ModelHDS}
The Case Study \ref{CS:Chronic} text is interpreted so as to identify the healthcare delivery system processes and resources.  They are modeled at a level of aggregation typical of clinical narratives.  For example, the narrative states `... a parietoccipital surgical approach was performed' \cite{Park2005}.  Here, the healthcare delivery system process is aggregated to `Perform surgery' and the `neurosurgery' resources describes the aggregation of human and technical resources including the neurosurgery team, room and equipment.  Second, an `outside clinic' resource is added to reflect the individual entering and exiting the healthcare delivery system in the case's six clinical visits.  As discussed in Section \ref{Sec:ChronicCare}, the transportation capabilities within the clinic are assumed to be always available, of relatively short duration, and of sufficient capacity.   They are eliminated from the knowledge base so as to focus on the more valuable healthcare delivery capabilities of transformation, decision and measurement.  Therefore, the transportation processes are reduced to `Enter clinic' and `Exit clinic'.  Furthermore, given the long term nature of this example, the decision processes of care planning and care scheduling are combined.  

\vspace{-0.1in}
 \begin{figure}[h]
  \begin{center} 
   \includegraphics[scale=0.485, trim=1.66cm 15.98cm 1.39cm 1.79cm, clip=true]{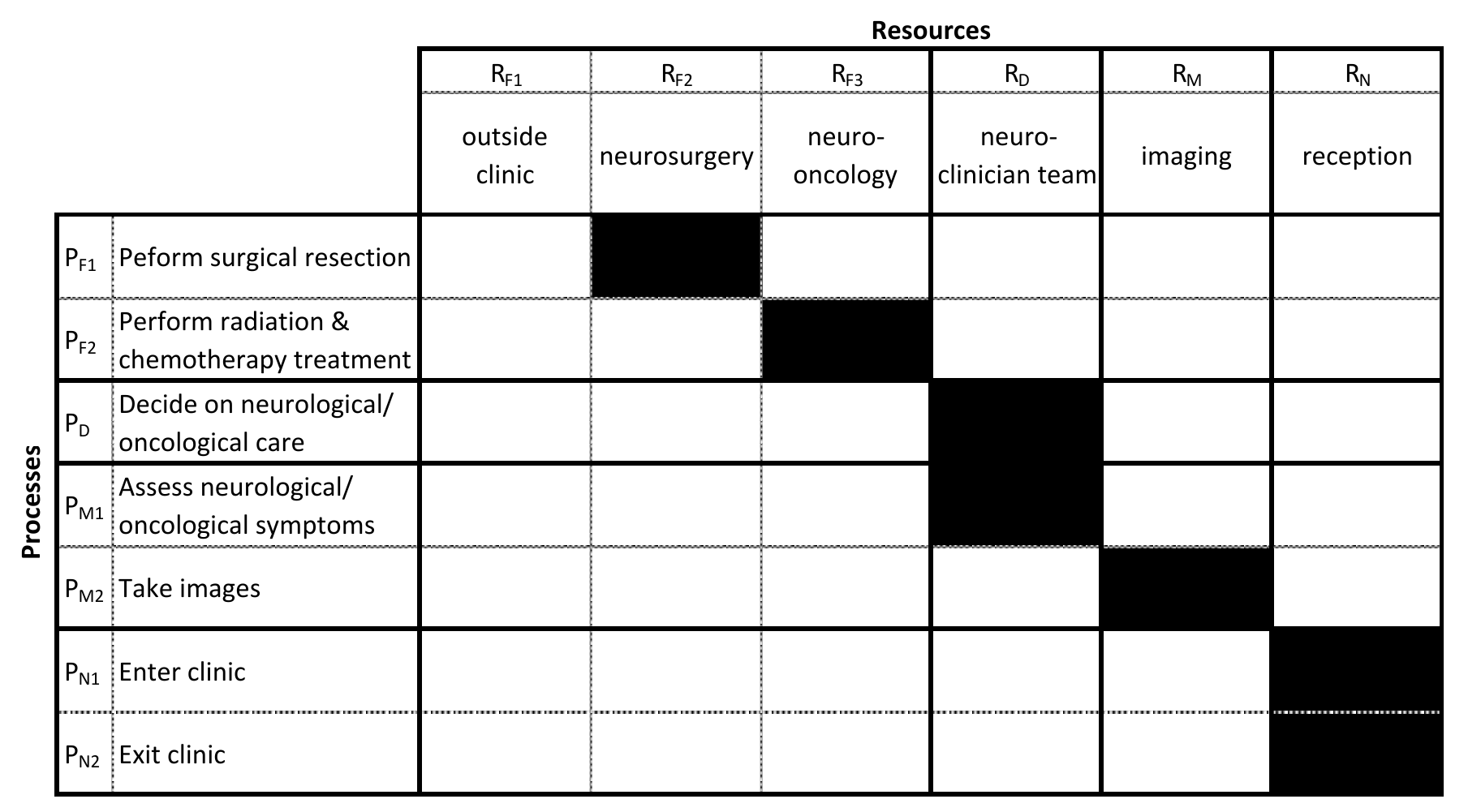} 
         \vspace{-0.26in}
    \caption{\textbf{Chronic Care Healthcare Delivery System Knowledge Base} $J_{S}$ with allocated processes to resources (dark filled).}
   \label{Fig:JS}
   \end{center}
   \vspace{-0.2in}
\end{figure}

\vspace{0.05in}
The resources and processes are classified as either transformation, decision, measurement or transportation and used to construct the system knowledge base $J_{S}$ shown in Figure \ref{Fig:JS}.  For simplicity, the system is assumed to not have any event constraints, $K_S=0$.  The system availability matrix and consequently the structural degrees of freedom can be calculated using Equation \ref{Eq:DOFS} such that $DOF_S=7$.

\begin{figure}[h]
       \vspace{-0.1in}
  \begin{center} 
   \includegraphics[width=3.in]{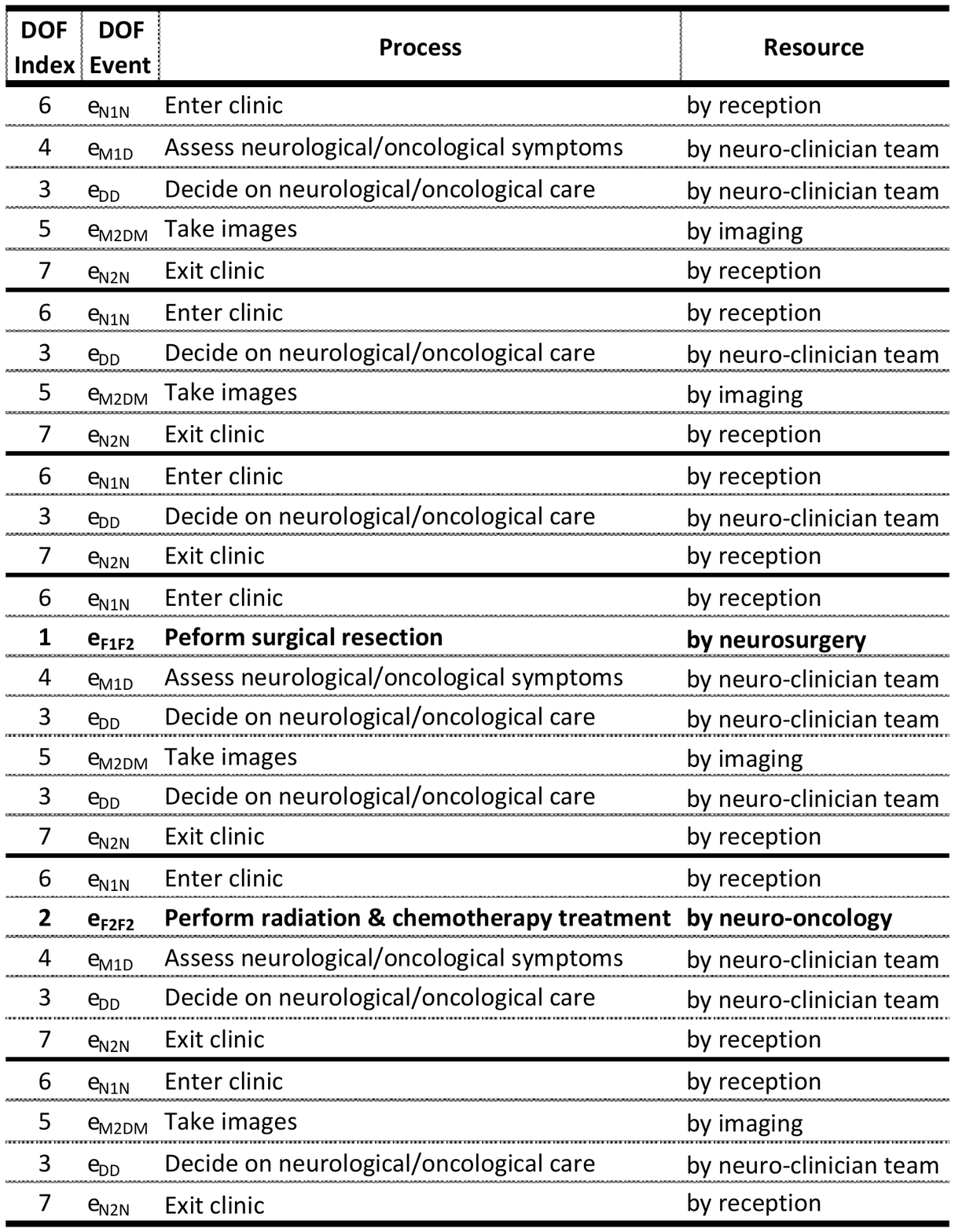} 
      \vspace{-0.1in}
    \caption{\textbf{Chronic Care Healthcare Delivery System Events} based on the Neuro-Oncology Case narrative in terms of the Healthcare Delivery System Events found in $J_{S}$.  The healthcare delivery transformational process events are in bold.}
   \label{Fig:EventList}
   \end{center}
   \vspace{-0.2in}
\end{figure}

The Case Study \ref{CS:Chronic} narrative is then rewritten as a string of healthcare delivery system events ${\cal E}_S$ as shown in Figure \ref{Fig:EventList}.  Each event in ${\cal E}_S$ has a unique index and its associated combination of process and resource.  The transformational events are highlighted in bold.  These events are effectively an \emph{untimed} scheduled events list and are used to generate the healthcare delivery system Petri net firing vectors.  

\begin{figure}[h]
  \begin{center} 
   \includegraphics[scale=0.48, trim=0.009cm 0.0095cm 0.062cm 0.0532cm, clip=true]{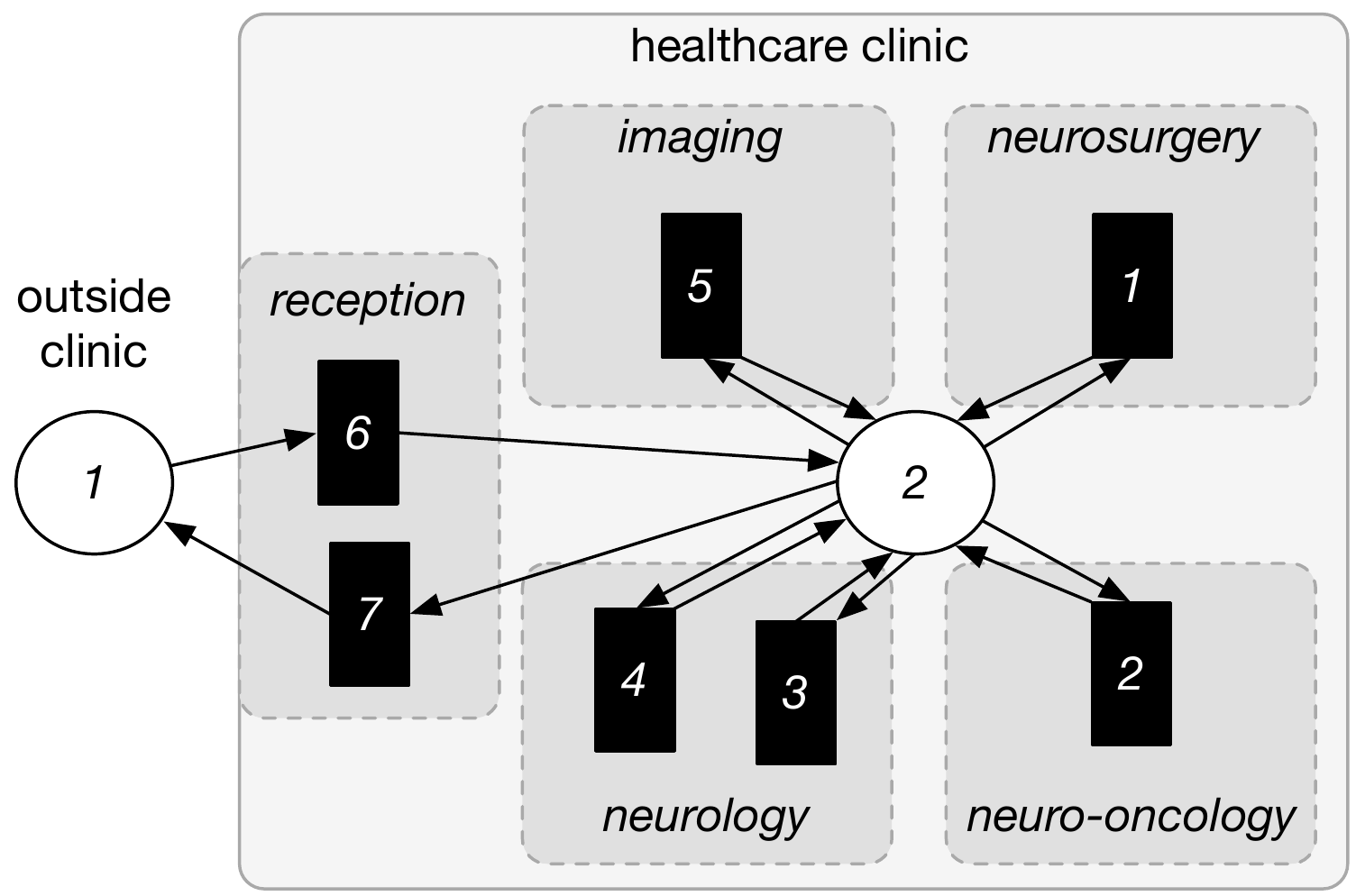} 
      \vspace{-0.126in}
    \caption{\textbf{Chronic Care Healthcare Delivery System Petri net}.  The places (\emph{circles}), represent  1=`outside clinic' and 2=`healthcare clinic'.  The transitions (\emph{rectangles}), are numbered based on the healthcare delivery system structural degrees of freedom index $\psi_i$.}
   \label{Fig:HDSnet}
   \end{center}
\vspace{-0.2in}
\end{figure}

Finally, the healthcare delivery system Petri net is constructed.  As discussed in Section \ref{Sec:ChronicCare}, the chronic care abstraction is utilized to abstract the many healthcare delivery system resources to a single `healthcare clinic' resource.  Figure \ref{Fig:HDSnet} shows the Petri net of the healthcare delivery system superimposed on a light-grey physical layout.  

\vspace{-0.1in}
 \begin{figure}[h]
  \begin{center} 
   \includegraphics[width=3.25in, scale=0.335, trim=0.cm 0.0cm 0.02cm 0.0cm, clip=false]{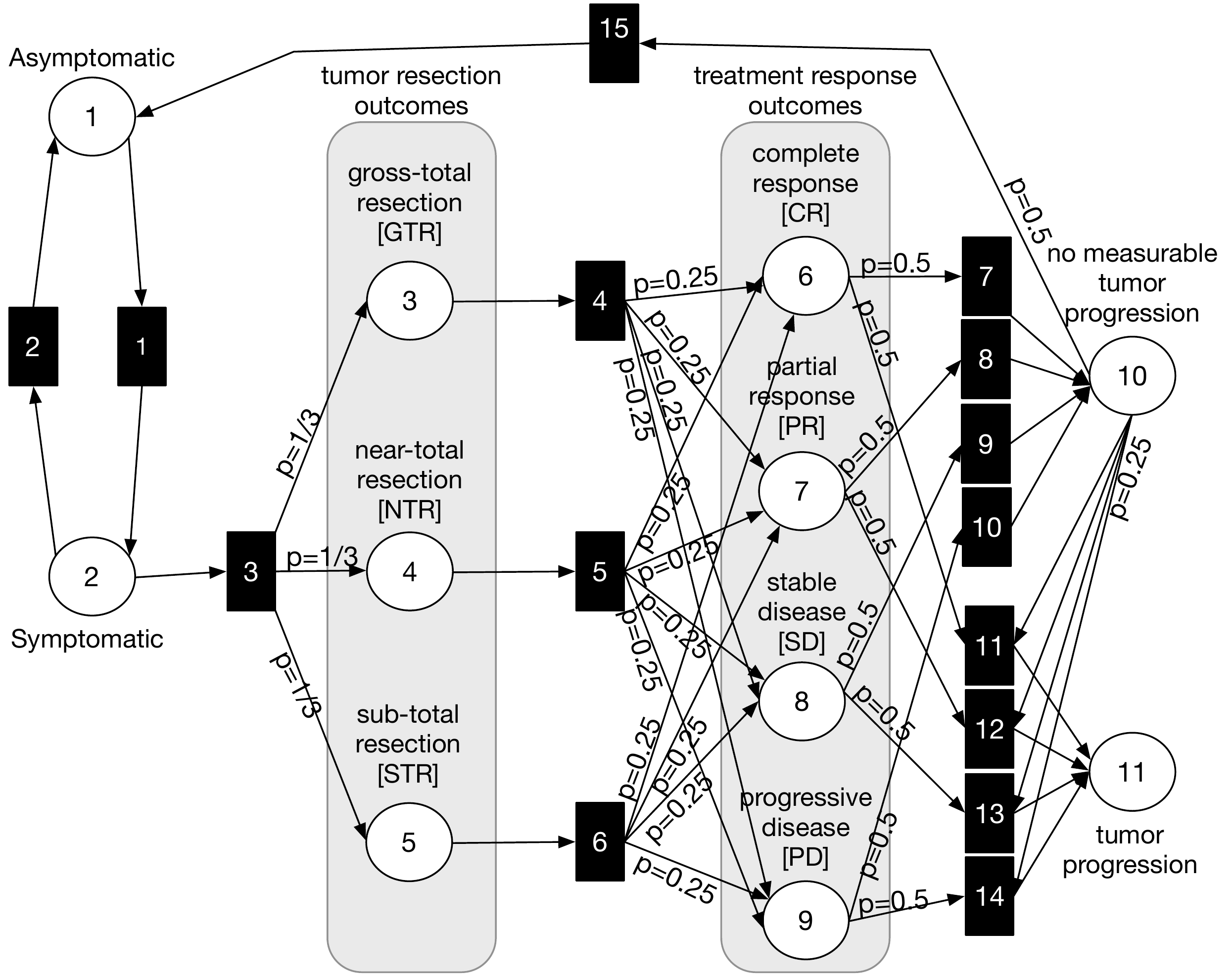}  
   \vspace{-0.14in}
    \caption{\textbf{Chronic Care Individual Net}: visualizes the health states at the places \emph{(circles)} and the health events at the transitions \emph{(rectangles)} representing the stochastic human process or the healthcare delivery system transformational process firing.} 
   \label{Fig:IndividualNet}
   \end{center}
   \vspace{-0.25in}
\end{figure}

\vspace{-0.12in}
\subsection{Modeling the Individual Health Net of the Neuro-Oncology Case} \label{Sec:Modelindividual}
The Case Study \ref{CS:Chronic} narrative and its associated healthcare delivery system events now serve to determine the health net for the neuro-oncology patient as shown in Figure \ref{Fig:IndividualNet}.  The health states are identified based on typical clinical outcomes of the healthcare transformational processes:  `Perform surgical resection' and `Perform radiation \& chemotherapy treatment'.

Because the health net is fuzzy, the outcome of any given transition is probabilistic.   For example, after a tumor resection three different outcomes are possible to describe the extent of resection:  gross-total resection [GTR], near-total resection [NTR] and sub-total resection [STR]).  For simplicity, the probabilities of these outcomes are assumed to be equal and in practice would be validated with clinical data of surgery outcomes.  Next, three transitions occur in parallel to represent radiation \& chemotherapy of the tumor/cancer in its current condition.   This leads to four possible treatment response outcomes based on the McDonald criterion \cite{Wen2010} (i.e. complete response [CR], partial response [PR], stable disease [SD], progressive disease [PD]).  These lead to states associated with further tumor progression, or no measurable tumor progression followed by an asymptomatic state.  In this case, the health net, as a fuzzy-timed Petri net, not only shows the dynamic evolution of an individual's distributed health state in the clinical sense but it's stochastic nature lends itself to Bayesian statistics.

\vspace{-0.25in}
\subsection{Modeling the Coordination of the Healthcare Delivery System Petri Net \& Individual Health Net of the Neuro-Oncology Case}
\label{Sec:Dyn} 

\begin{figure}[h]
  \begin{center} 
   \includegraphics[scale=0.34985, ]{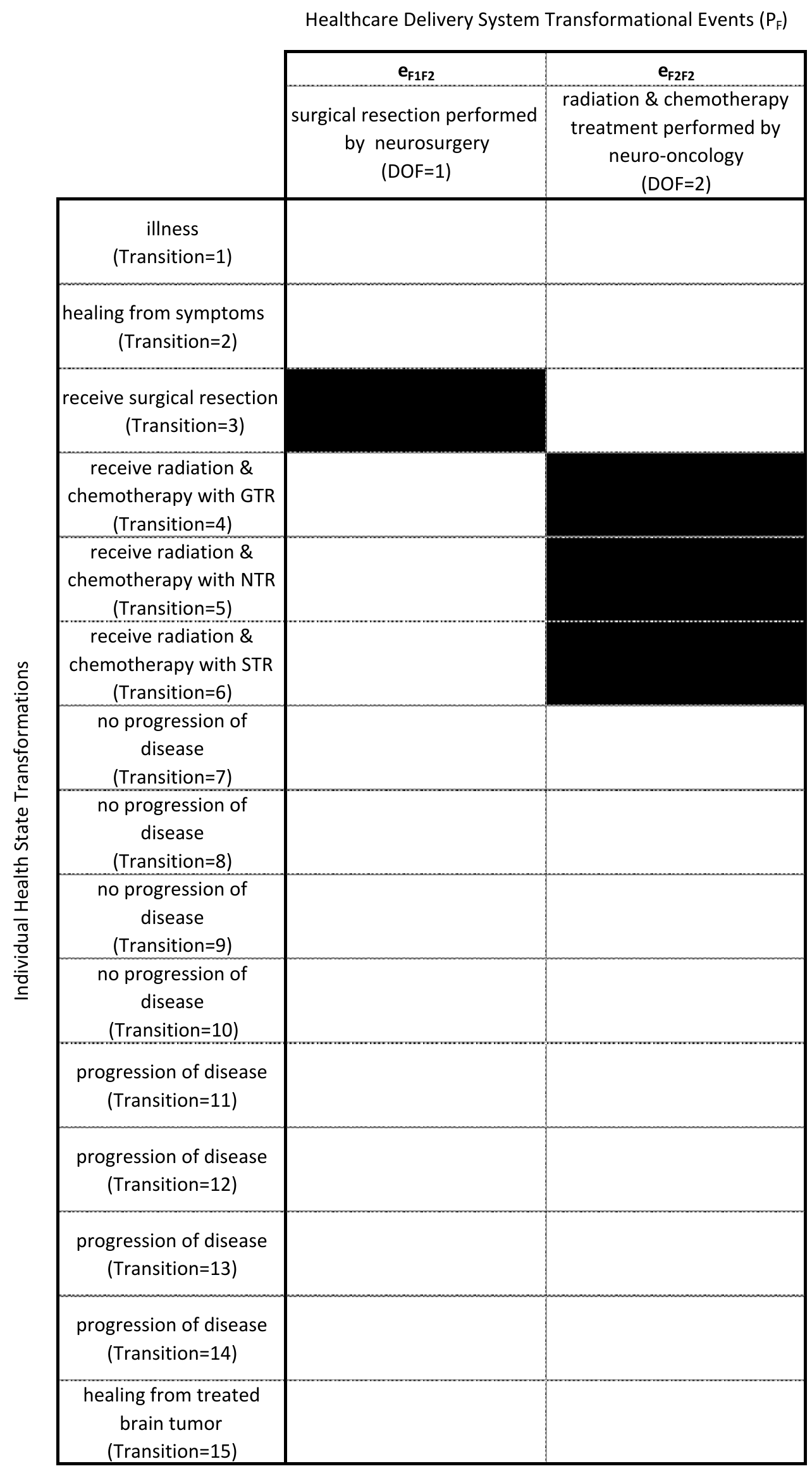} 
      \vspace{-0.1in}
    \caption{\textbf{Chronic Care Individual Transformation Feasibility Matrix} $\Lambda_{F}$.}
   \label{Fig:FeasibilityMatrix}
   \end{center}
      \vspace{-0.2in}
\end{figure}

To complete the model, the healthcare delivery system Petri net and the individual health net must be coordinated.  The individual transformation feasibility matrix, shown in Figure \ref{Fig:FeasibilityMatrix}, is constructed by linking the individual health net transitions (i.e. health events) to the corresponding healthcare delivery system transformational events (i.e. transformation process $P_F$).  Note that the radiation and chemotherapy transformation process is tied to three health events; not just one.  This is because the individual's health evolves differently to the same stimulus depending on their current condition.    
 \vspace{-0.1in}

\begin{figure}[]
  \begin{center} 
      \includegraphics[width=3.5in]{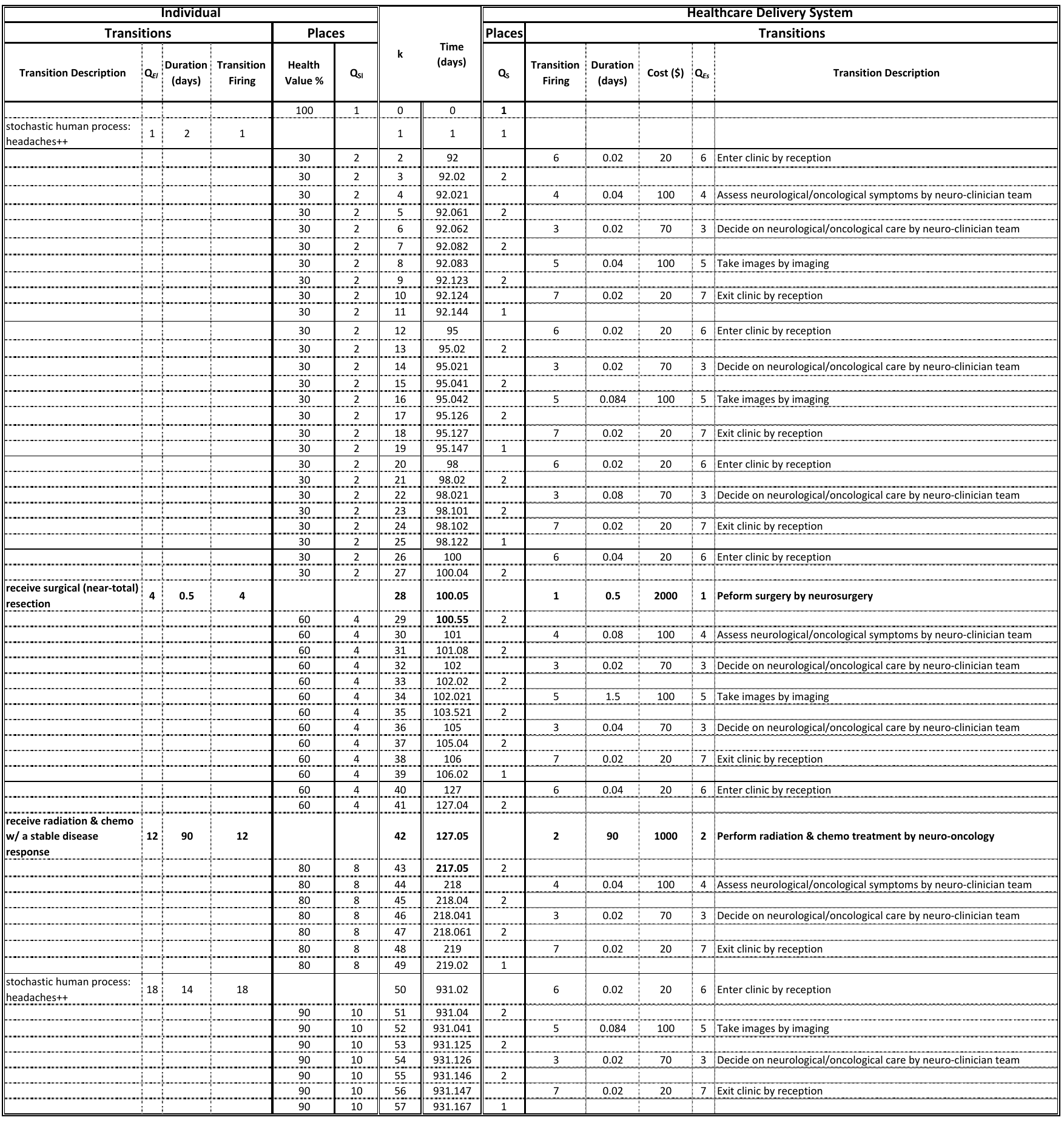} 
     \vspace{-0.25in}
    \caption{\textbf{Chronic Care Dynamics of the Two Petri nets}:  the healthcare delivery system Petri net and the individual net are now synchronized.  The scheduled events of each Petri net are shown side by side.}
   \label{Fig:DynamicTable}
   \end{center}
   \vspace{-0.2in}
\end{figure}

The synchronized dynamics of the healthcare delivery system Petri net and the individual net are shown in Figure \ref{Fig:DynamicTable} as two scheduled event lists side by side.

{\color{black} Finally, the healthcare delivery system operating cost and individual health outcome dynamics for this Neuro-Oncology Chronic case can be shown over time (k) in Figure \ref{Fig:Cost_HealthOutcome2}.

 \begin{figure}[h]
  \begin{center}   
   \includegraphics[width=3in]{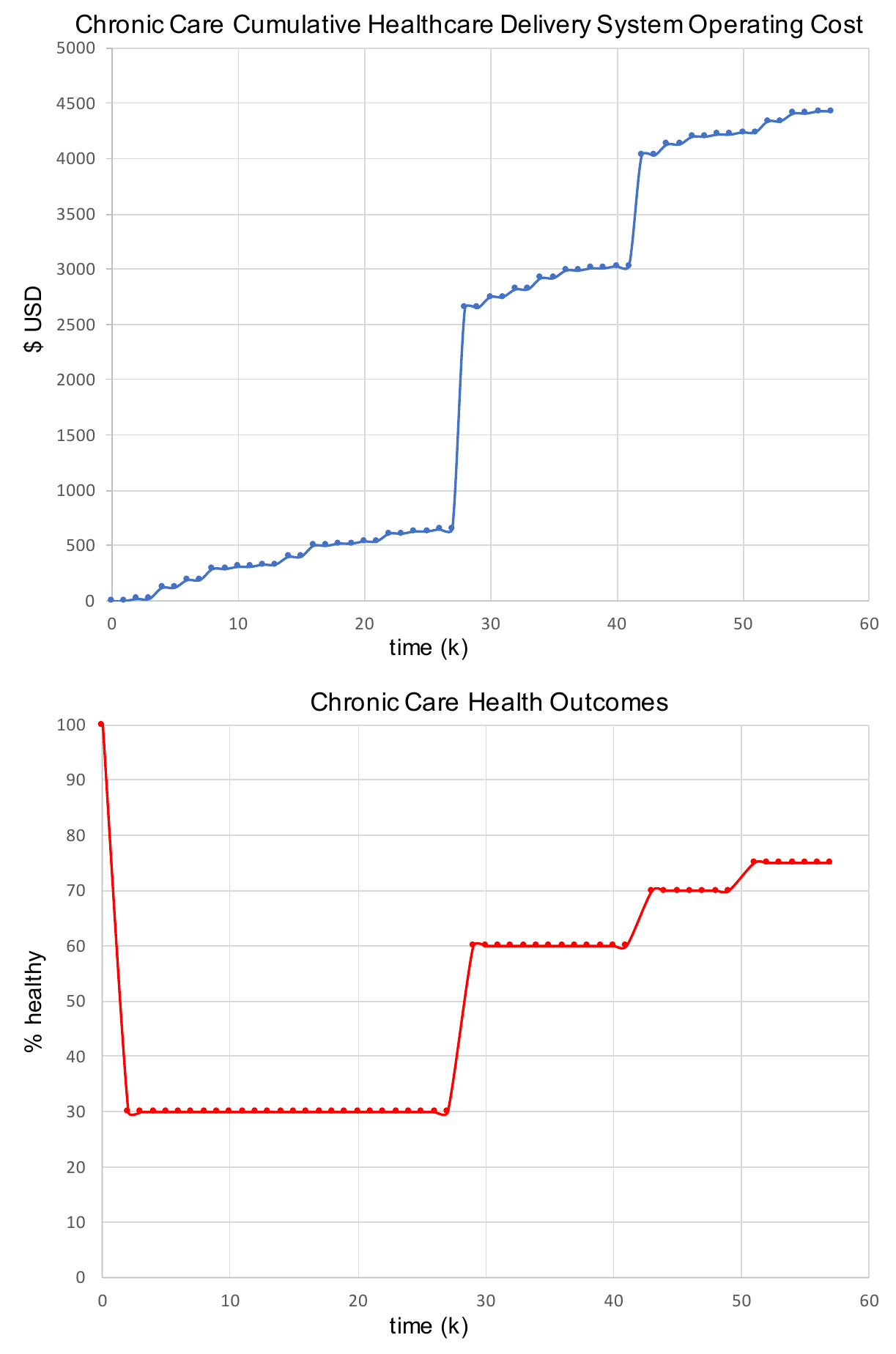} 
   \vspace{-0.2in}
    \caption{\textbf{Chronic Care Healthcare Delivery System Operating Cost in US dollars (top-blue) and Individual Health Outcome as \% healthy (bottom-red).}}
    \label{Fig:Cost_HealthOutcome2}
   \end{center}
   \vspace{-0.25in}
\end{figure}
}

\vspace{-0.2in}
\section{Results \& Discussion} \label{Sec:Discussion}

The illustrative examples demonstrate the dynamic model for personalized healthcare delivery and managed health outcomes.  Such a model has several important aspects.  It applies equally to the care of both acute and chronic conditions, it transparently describes health outcomes, and it transparently links cost and outcomes.

First, the two illustrative examples demonstrate the model's applicability to both acute and chronic care.  The acute care dynamic modeling resembles those commonly found in industrial engineering and operations research literature \cite{Gunal2010, Jacobson2006, Turkcan2012, Alvarado2012}.  It emphasizes the importance of scheduling and minimized queuing.  At the timescale of a clinic visit, acute care decision processes like care planning and scheduling are critical.   The timeliness of decision-making was highlighted in the acute care of the orthopedic case.  Acute care requires a more granular resolution of healthcare delivery system capabilities (i.e. structural degrees of freedom).  Consequently, many more are utilized per visit relative to chronic care.   The utilization of spatially distributed transformative, decision, and measurement capabilities within a short period of time naturally raises questions of transportation (e.g. in emergency rooms) and queues (e.g. in patient care).  In chronic care, these concerns are diminished.  The model abstracts away transportation so as to focus on the coordination of transformation, decision and measurement processes.  

The health net in this model is an important contribution that serves to transparently describe health outcomes.  In acute care, the health net tends to cycle back to an initially healthy state in a fairly short period of time; and perhaps within a single visit.  In chronic care, not only are multiple clinical visits required but the state of the individual's health must be tracked in the meantime.  While in some chronic conditions a return to a healthy state is possible, in most instances the healthcare delivery system must actively track and manage its degradation.   

Indeed, the most important aspect of the model is its coherence between the healthcare delivery system and the individual's health state.  The states of the Petri nets are tied directly and should ideally be coordinated in order to deliver effective care.  Whether for acute or chronic conditions, time is of the essence.   Because the health nets have stochastic processes that will fire spontaneously, the healthcare delivery system must take timely and coordinated action to avoid adverse and negative health outcomes.   

Finally, it is important to recognize that each healthcare delivery system degree of freedom incurs a cost every time it is fired.  Therefore, as the two Petri nets evolve simultaneously, the discrete event simulation transparently reveals the accumulation of incurred cost versus the evolution of health outcomes   {\color{black} as shown in Figures \ref{Fig:Cost_HealthOutcome} and \ref{Fig:Cost_HealthOutcome2}.}

\vspace{-0.1in}
\section{Conclusion \& Future Work} \label{Sec:Conclusion}

In conclusion, this paper develops the dynamic system model for personalized healthcare delivery and managed individual health outcomes.  The dynamics of the model rests upon the developed systems architecture from prior work.  This work draws upon a hetero-functional graph theory rooted in Axiomatic Design for Large Flexible Engineering Systems and Petri nets.   
The dynamic model coordinates the healthcare delivery system and individual net.  The healthcare delivery net evolves as the transitions fire when the system is utilized, while the individual net evolves as the individual's health state evolves due to the spontaneous firing of stochastic process and as the individual receives transformative processes by the healthcare delivery system.    The dynamic model was then demonstrated for two illustrative examples: an acute care and chronic care.  The contrast of the two examples shows the versatility of the model to handle the queueing needs of acute care and the coordination needs of chronic care.  {\color{black} Furthermore, the dynamic model was used to produce two parametric functions of time dynamically showing healthcare delivery system operating cost over time and patient outcome values over time.  These could be used to understand what healthcare capability utilizations or utilization patterns lead to better patient outcome values.}  

The development of the model opens several new avenues for future work.   The Petri net firing vectors indicated as inputs to the model provide an opportunity for the development of rigorous decision-making algorithms.  Second, the model may be applied to new case studies of potentially larger scope.   The clear trade-offs between cost and health outcomes is likely to be of interest to many healthcare delivery system stakeholders including clinicians, healthcare facility administrators, insurance companies, and regulators.  
Finally, as the need for such a model matures, new approaches to automated model building that integrate with healthcare enterprise information systems or capture health state information through Internet-of-Things enabled sensor devices is likely to grow.

\bibliographystyle{IEEEtran}
\bibliography{SHPublications,librarySystemModel,Library}

\end{document}